\documentclass[twocolumn]{aastex7}

\usepackage{xcolor,tikz}
\usepackage{gensymb}
\usepackage{multirow}
\usepackage{subcaption}
\begin{document}

\title{The ZZ Ceti Instability Strip for The Most Massive White Dwarf Pulsators}

\author[0009-0009-9105-7865]{Gracyn Jewett}
\affiliation{Homer L. Dodge Department of Physics and Astronomy, University of Oklahoma, 440 W. Brooks St., Norman OK, 73019, USA}
\email[show]{gjewett@ou.edu}  

\author[0000-0001-6098-2235]{Mukremin Kilic} 
\affiliation{Homer L. Dodge Department of Physics and Astronomy, University of Oklahoma, 440 W. Brooks St., Norman OK, 73019, USA}
\email{kilic@ou.edu}

\author[0000-0001-7143-0890]{Adam Moss} 
\affiliation{Department of Astronomy, University of Florida, Bryant Space Science Center, Stadium Road, Gainesville, FL 32611, USA}
\email{Adam.G.Moss-1@ou.edu}

\author[0000-0002-0006-9900]{Alejandro H. Córsico}
\affiliation{Grupo de Evolución Estelar y Pulsaciones. Facultad de Ciencias Astronómicas y Geofísicas, Universidad Nacional de La Plata, Paseo del
Bosque s/n, (1900) La Plata, Argentina}
\affiliation{Instituto de Astrofísica de La Plata, IALP (CCT La Plata), CONICET-UNLP, Argentina}
\email{alejandrocorsico@gmail.com}

\author[0000-0002-0948-4801]{Matthew J. Green}
\affiliation{Homer L. Dodge Department of Physics and Astronomy, University of Oklahoma, 440 W. Brooks St., Norman OK, 73019, USA}
\email{matthew.j.green-2@ou.edu}

\author[0000-0002-6603-994X]{Murat Uzundag}
\affiliation{Institute of Astronomy, KU Leuven, Celestijnenlaan 200D, 3001, Leuven, Belgium}
\email{muratuzundag.astro@gmail.com}

\author[0000-0003-2368-345X]{Pierre Bergeron} 
\affiliation{Département de Physique, Université de Montréal, C.P. 6128, Succ. Centre-Ville, Montréal, QC H3C 3J7, Canada}
\email{pierre.bergeron.1@umontreal.ca}

\author[0000-0002-4462-2341]{Warren R.\ Brown}
\affiliation{Center for Astrophysics, Harvard \& Smithsonian, 60 Garden Street, Cambridge, MA 02138, USA}
\email{wbrown@cfa.harvard.edu}

\author[0000-0002-8079-0772]{Francisco C. De Gerónimo}
\affiliation{Grupo de Evolución Estelar y Pulsaciones. Facultad de Ciencias Astronómicas y Geofísicas, Universidad Nacional de La Plata, Paseo del
Bosque s/n, (1900) La Plata, Argentina}
\affiliation{Instituto de Astrofísica de La Plata, IALP (CCT La Plata), CONICET-UNLP, Argentina}
\email{degeronimofrancisco@gmail.com}

\author[0000-0002-6153-7173]{Alberto Rebassa-Mansergas}
\affiliation{Departament de Física, Universitat Politècnica de Catalunya, c/Esteve Terrades 5, 08860 Castelldefels, Spain}
\affiliation{Institut d'Estudis Espacials de Catalunya, Esteve Terradas, 1, Edifici RDIT, Campus PMT-UPC, 08860 Castelldefels, Barcelona, Spain}
\email{alberto.rebassa@upc.edu}

\author[0000-0002-3316-7240]{Alex J. Brown}
\affiliation{Hamburger Sternwarte, University of Hamburg, Gojenbergsweg 112, 21029 Hamburg, Germany}
\email{alexjbrown.astro@gmail.com}

\author[0000-0003-4236-9642]{Vikram S. Dhillon}
\affiliation{Astrophysics Research Cluster, School of Mathematical and Physical Sciences, University of Sheffield, Sheffield, S3 7RH, UK}
\affiliation{Instituto de Astrof\'{\i}sica de Canarias, E-38205 La Laguna, Tenerife, Spain}
\email{Vik.Dhillon@sheffield.ac.uk}

\author[0000-0001-7221-855X]{Stuart Littlefair}
\affiliation{Astrophysics Research Cluster, School of Mathematical and Physical Sciences, University of Sheffield, Sheffield, S3 7RH, UK}
\email{s.littlefair@sheffield.ac.uk}

\begin{abstract}
%250 word limit 
We present time-series photometry of 31 massive DA white dwarfs with $M\gtrsim 0.9~M_\odot$ within the ZZ Ceti instability strip from the Montreal White Dwarf Database 100 pc
sample. The majority of the targets had no previous time-series photometry available, though several were classified as non-variable or potential pulsators
in the literature. Out of the 31 candidates, we confirm 16 as pulsating. %and do not detect pulsations in 15. 
Our observations at three observatories have led us to discover
the most massive pulsating white dwarf currently known, J0959$-$1828 ($M=1.32$ or $1.27~M_\odot$ for a CO versus ONe core), which is slightly more massive than the
previous record holder J0049$-$2525. We study the sample properties of massive ZZ Ceti white dwarfs, and find several trends with their weighted mean
periods. As predicted by theory, we see an increase in the weighted mean periods with decreasing effective temperature, and a decrease in pulsation
amplitudes at the red edge of the instability strip. Furthermore, the weighted mean periods decrease with increasing stellar mass.  Our observations show
that the ZZ Ceti instability strip may not be pure at high masses. This is likely because the non-variable white dwarfs in the middle of the strip may be
weakly magnetic, which could escape detection in the available low-resolution spectroscopy data, but may be sufficient to suppress pulsations. Extensive
follow-up observations of the most massive white dwarfs in our sample have the potential to probe the interior structures and core-compositions of these
white dwarfs with significantly crystallized cores.
\end{abstract}


\section{Introduction} 

\begin{deluxetable*}{crrrrccccccc}
\tabletypesize{\tiny}
\tablecolumns{12} \tablewidth{0pt}
\caption{Observational and physical parameters of the ZZ Ceti candidates observed. Mass and cooling age estimates are based on CO core models \citep{jewett2024}.}
\label{tab:obsphys}
\tablehead{\colhead{Object name} & \colhead{Gaia DR3 ID} & \colhead{RA} & \colhead{DEC} & \colhead{Parallax} & \colhead{G} & \colhead{$G_{BP}$} & \colhead{$G_{RP}$} & \colhead{log (g)}& \colhead{$T_{\rm eff}$} & \colhead{Mass} & \colhead{Cooling Age}\\
 & &  ($^{\circ}$) & ($^{\circ}$) & (mas) & (mag) & (mag) & (mag) & (cgs) &(K) & $M_\odot$ & (Gyr)}
\startdata
J0039$-$0357 & 2527618112309283456 & 9.78326 & $-$3.95607 & 11.30 $\pm{0.22}$ &  18.61 & 18.62 & 18.67 & 9.20 &11871 $\pm$214 & 1.271 $\pm$0.009 & 2.09 $\pm$0.06\\
J0050$-$2826 & 2342438501397962112 & 12.71700 & $-$28.43495 & 11.07 $\pm{0.11}$ & 17.81 & 17.86 & 17.80 &   8.73&11320 $\pm$155 & 1.061 $\pm$0.011 & 1.72 $\pm$0.08\\
J0127$-$2436 & 5040290528701395456 & 21.89464 & $-$24.60557 & 12.36 $\pm{0.19}$ &  18.69 & 18.74 & 18.69 &   9.24 & 11236 $\pm$214 & 1.284 $\pm$0.009 & 2.22 $\pm$0.06\\
%J0135+5722 & 412839403319209600 & 23.82370  & 57.37989 & 19.66 $\pm{0.05}$ & 16.66 & 16.67 & 16.71 & 8.90 & 12415 $\pm{87}$ & 1.153 $\pm{0.004}$ & 1.75 $\pm{0.03}$ \\
J0154$+$4700 & 356186555597277440 & 28.64365 & 47.01313 & 10.21 $\pm{0.12}$ &  17.93 & 17.95 & 17.96 &    8.76& 11838 $\pm$235 & 1.077 $\pm$0.014 & 1.59 $\pm$0.10\\
J0158$-$2503 & 5121833510769131136 & 29.69913 & $-$25.05308 & 11.21 $\pm{0.10}$ & 17.79 & 17.79 & 17.80 &  8.84& 12234 $\pm$94 & 1.122 $\pm$0.007 & 1.68 $\pm$0.04\\
J0204$+$8713 & 575585919005741184 & 31.12914 & 87.22579 & 11.11 $\pm{0.08}$ & 17.79 & 17.85 & 17.78 & 8.71 & 11135 $\pm$207& 1.053 $\pm$0.015 & 1.75 $\pm$0.11\\
J0408$+$2323 & 53716846734195328 & 62.01259 & 23.39512 & 12.40 $\pm{0.09}$ & 17.29 & 17.32 & 17.30 &  8.66& 12053 $\pm$110 & 1.024 $\pm$0.008 & 1.22 $\pm$0.06\\
J0538$+$3212 & 3447991090873280000 & 84.74184 & 32.20788 & 10.30 $\pm{0.10}$ & 17.51 & 17.53 & 17.57 &   8.61& 12454 $\pm$154 & 0.994 $\pm$0.010 & 0.95 $\pm$0.05 \\
J0634$+$3848 & 945007674022721280 & 98.56907 & 38.81528 & 21.88 $\pm{0.04}$ & 15.73 & 15.74 & 15.75 &   8.50& 12210 $\pm$106 & 0.926 $\pm$0.006 & 0.81 $\pm$0.02 \\
J0657$+$7341 & 1114813977776610944 & 104.29631 & 73.69572 & 11.86 $\pm{0.09}$ &  17.65 & 17.67 & 17.68 &   8.86& 12609 $\pm$119 & 1.134 $\pm$0.007 & 1.61 $\pm$0.05\\
J0712$-$1815 & 2934281803636268416 & 108.08535 & $-$18.26537 & 17.87 $\pm{0.05}$ & 16.40 & 16.41 & 16.41 &   8.59& 11742 $\pm$136 & 0.980 $\pm$0.008 & 1.09 $\pm$0.05\\
J0725$+$0411 & 3139633462883694976 & 111.45264 & 4.19282 & 11.37 $\pm{0.08}$ &  17.22 & 17.23 & 17.26 &   8.53& 12022 $\pm$88 & 0.940 $\pm$0.008 & 0.88 $\pm$0.02 \\
J0912$-$2642 & 5649808720867457664 & 138.11673 & $-$26.70103 & 27.48 $\pm{0.04}$ & 16.41 & 16.40 & 16.46 &   9.17& 12973 $\pm$115 & 1.262 $\pm$0.002 & 1.84 $\pm$0.03\\
J0949$-$0730 & 3819743428284589696 & 147.41109 & $-$7.501786 & 12.42 $\pm{0.08}$ &  17.32 & 17.33 & 17.39 &   8.77& 12941 $\pm$184 & 1.084 $\pm$0.008 & 1.26 $\pm$0.07\\
J0950$-$2841 & 5464929134894103808 & 147.74005 & $-$28.68750 & 12.76 $\pm{0.09}$ &  17.34 & 17.31 & 17.41 &  8.84& 13335 $\pm$203 & 1.124 $\pm$0.006 & 1.35 $\pm$0.06\\
J0959$-$1828 & 5671878015177884032 & 149.88858 & $-$18.47373 & 12.76 $\pm{0.09}$ &  18.13 & 18.15 & 18.14 & 9.38& 11995 $\pm$176 & 1.320 $\pm$0.004 & 1.83 $\pm$0.05 \\
J1052$+$1610 & 3981634249048141440 & 163.15010 & 16.17275 & 13.39 $\pm{0.12}$ &  17.27 & 17.29 & 17.26 &  8.66& 11256 $\pm$84 & 1.020 $\pm$0.009 & 1.52 $\pm$0.06\\
J1106$+$1802 & 3983606596814071680 & 166.51878  & 18.04182 & 12.33 $\pm$0.09 & 17.51 & 17.51 & 17.55 & 8.86 & 12877 $\pm$269 & 1.131 $\pm$0.011 & 1.51 $\pm$0.09 \\
J1243$+$4805 & 1543370904111505408 & 190.92339 & 48.09304 & 12.51 $\pm{0.05}$ &  16.97 & 16.96 & 17.02 &  8.57& 12751 $\pm$101 & 0.966 $\pm$0.006 & 0.81 $\pm$0.02\\
J1342$-$1413 & 3606080968656361344 & 205.67123 & $-$14.22819 & 20.54 $\pm{0.04}$ & 15.98 & 16.00 & 15.96 &8.47 & 11351 $\pm$66 & 0.902 $\pm$0.006 & 0.93 $\pm$0.02\\
J1451$-$2502 & 6229330032504603136 & 222.96896 & $-$25.04411 & 12.63 $\pm{0.15}$ & 17.39 & 17.43 & 17.37 &  8.65& 11189 $\pm$97 & 1.016 $\pm$0.011 & 1.53 $\pm$0.07\\
J1552$+$0039 & 4410623858974488832 & 238.15940 & 0.65268 & 9.72 $\pm{0.25}$ & 18.49 & 18.48 & 18.55 &  9.12& 13508 $\pm$233 & 1.245 $\pm$0.011 & 1.70 $\pm$0.05\\
J1626$+$2533 & 1304081783374935680 & 246.74813 & 25.55767 & 11.88 $\pm{0.07}$ &  17.59 & 17.60 & 17.66 &  8.87& 13202 $\pm$189 & 1.140 $\pm$0.007 & 1.46 $\pm$0.06\\
J1656$+$5719 & 1433629812475800320 & 254.16291 & 57.31760 & 12.69 $\pm{0.07}$ &  17.40 & 17.43 & 17.41 &  8.70& 11551 $\pm$143 & 1.047 $\pm$0.010 & 1.54 $\pm$0.07\\
J1722$+$3958 & 1346883876962000000 & 260.66923 & 39.96965 & 10.30 $\pm{0.07}$ &  17.81 & 17.82 & 17.78 & 8.62& 11069 $\pm$109 & 0.997 $\pm$0.010 & 1.45 $\pm$0.07 \\
J1819$+$1225 & 4484736543328721792 & 274.85822 & 12.43255 & 9.77 $\pm{0.13}$ &  18.13 & 18.13 & 18.11 &  8.83& 11929 $\pm$185 & 1.116 $\pm$0.012 & 1.76 $\pm$0.08 \\
J1928$+$1526 & 4321498378443922816 & 292.06065 & 15.44412 & 10.26 $\pm{0.10}$ &  17.73 & 17.75 & 17.73 &  8.64& 11621 $\pm$141 & 1.011 $\pm$0.012 & 1.31 $\pm$0.08\\
J1929$-$2926 & 6764486080026955136 & 292.30264 & $-$29.44520 & 11.28 $\pm{0.16}$ &  17.69 & 17.71 & 17.71 &  8.72& 11455 $\pm$220 & 1.054 $\pm$0.016 & 1.62 $\pm$0.12\\
J2026$-$2254 & 6849850998873128704 & 306.69792 & $-$22.91429 & 14.20 $\pm{0.08}$ &  16.79 & 16.80 & 16.79  &  8.49& 11405 $\pm$88 & 0.917 $\pm$0.008 & 0.95 $\pm$0.03\\
J2107$+$7831 & 2284856630775525376 & 316.96357 & 78.53158 & 10.15 $\pm{0.08}$ &  17.83 & 17.82 & 17.92 &8.79 & 12707 $\pm$202 & 1.099 $\pm$0.009 & 1.41 $\pm$0.08\\
J2208$+$2059 & 1781605382738862592 & 332.13117 & 20.98618 & 11.09 $\pm{0.09}$ & 17.48 & 17.55 & 17.46 & 8.53& 11091 $\pm$99 & 0.941 $\pm$0.011 & 1.12 $\pm$0.05\\
\enddata
\end{deluxetable*}

The vast majority of stars will evolve into white dwarfs \citep{fontaine01,althaus}, and the majority of white dwarfs appear as DA white dwarfs, which
have hydrogen dominated atmospheres. DA white dwarfs experience a phase of pulsations when they reach $\sim 12,000$ K. Pulsating DA white dwarfs (DAVs) are commonly
referred to as ZZ Cetis, named after the prototype of this class \citep{landolt}. 

ZZ Ceti white dwarfs are found in a narrow temperature range between $\approx$10,400-12,400 K with periods ranging from 100-1,400 seconds \citep{corsico}.
Pinpointing the exact boundaries of the instability strip is no small task \citep{mukadam04,gianninas05,castanheira10,vangrootel12}. An improved set of
empirical boundaries were found by \citet{vincent}, who took advantage of Gaia parallaxes to constrain masses and used the results from several
photometric surveys to define the boundaries of the instability strip (see their Figure 4).

The mass distribution of the white dwarfs in the solar neighborhood peaks at 0.6 $M_\odot$ with a broad shoulder and tail towards higher masses
\citep{kilic2020,kilic25,obrien}, hence massive ($M\gtrsim0.9~M_\odot$) and ultra-massive ($M\gtrsim 1.05~M_\odot$) pulsating white dwarfs are rare. BPM 37093 is an ultra-massive ZZ Ceti white dwarf
with a mass of $1.037 \pm 0.008~M_\odot$ \citep{obrien} that was the most massive pulsator known at the time of its discovery
\citep{kanaan}. It is predicted that 90\% of its core is crystallized \citep{metcalfe}. \citet{hermes} targeted GD 518 based on 1D model atmosphere fits
to its optical spectrum, which indicated a mass of $1.20 \pm 0.03~M_{\odot}$ \citep{gianninas}. They detected three significant modes
of pulsations in this star with periods ranging from 425 to 595 s, making it the most massive pulsating white dwarf known at the time. 
A model fit including Gaia parallaxes and the photometric method \citep{bergeron19} results in a revised mass estimate of $1.114 \pm 0.006~M_{\odot}$
for GD 518 \citep{kilic25}. Additional discoveries of massive pulsating white dwarfs include SDSS J084021.26+522217.8 \citep{curd17}, which has
$M=0.98\pm 0.04~M_{\odot}$ \citep{vincent24}, and the most massive white dwarf known until now WDJ004917.14$-$252556.81 with $M=1.312 \pm 0.010~M_{\odot}$
\citep{kilic23b,2025ApJ...988...32C}.

\citet{jewett2024} presented a detailed model atmosphere analysis of all white dwarfs with $M>0.9~M_{\odot}$ and $T_{\rm eff}\geq11,000$ K in the
Montreal White Dwarf Database \citep[MWDD,][]{dufour17} 100 pc sample and the Pan-STARRS footprint. Because the temperature range of this sample of white dwarfs overlaps with the ZZ Ceti instability strip, they were able to identify massive DA white dwarfs that fell within the empirical boundaries of the strip. They found 8 previously known pulsating massive white dwarfs and 14 objects that were not observed to vary (NOV) in this sample.
They identified 22 ZZ Ceti candidates with no time-series follow-up at the time, including four objects with $M>1.2~M_{\odot}$. 
They also found several NOVs from \citet{vincent} that fall right in the middle of the ZZ Ceti strip, and identified J0135+5722 as one of the most
interesting in that sample with $M=1.15~M_\odot$. 
\citet{degeronimo}  have since confirmed multi-periodic pulsations in J0135+5722. 

In order to identify massive and ultra-massive pulsating white dwarfs and study the ZZ Ceti instability strip at the high mass end, we obtained time-series observations
of 31 of the candidates identified by \citet{jewett2024}. Here we present the light curves and Fourier transforms for all 31 systems,
and provide a list of the significant frequencies detected for each object. This study provides a detailed view of the ZZ Ceti instability strip
for the MWDD 100 pc sample of massive white dwarfs in the Pan-STARRS footprint. 

In Section 2, we discuss the sample of white dwarfs studied in this paper, details of our observations, and the analysis of our data. Section 3 discusses
the most massive pulsating white dwarfs in our sample in detail and presents the light curves and Fourier transforms for each star. Following this is a discussion of the non-variable
white dwarfs in Section 4. In Section 5, we explore trends in our data and discuss the purity of the ZZ Ceti instability strip for massive
white dwarfs. Finally, we make our concluding remarks in Section 6. 

\section{Observations and Analysis}

\subsection{Sample}

Table \ref{tab:obsphys} presents the observational and physical parameters of the 31 ZZ Ceti candidates from \citet{jewett2024}. 
The masses range from $0.90~M_{\odot}$ to $1.32~M_{\odot}$ assuming CO cores, including 22 candidates with $M>1~M_\odot$.  All but two of these targets have parallaxes $\geq10$ mas. The two targets with smaller parallaxes, J1552+0039 and J1819+1225, are part of the MWDD 100 pc sample
since that selection included objects with parallaxes consistent with being within 100 pc within $1\sigma$ based on Gaia DR2 astrometry.
Note that, time series photometry for one of the candidates in the \citet{jewett2024} sample, J0135+5722 (Gaia DR3 412839403319209600), was recently
presented by \citet{degeronimo}, hence we exclude it from this table. 

\begin{figure}
\includegraphics[width=3.5in, clip, trim={0.2in 0.1in 0.4in 0.4in}]{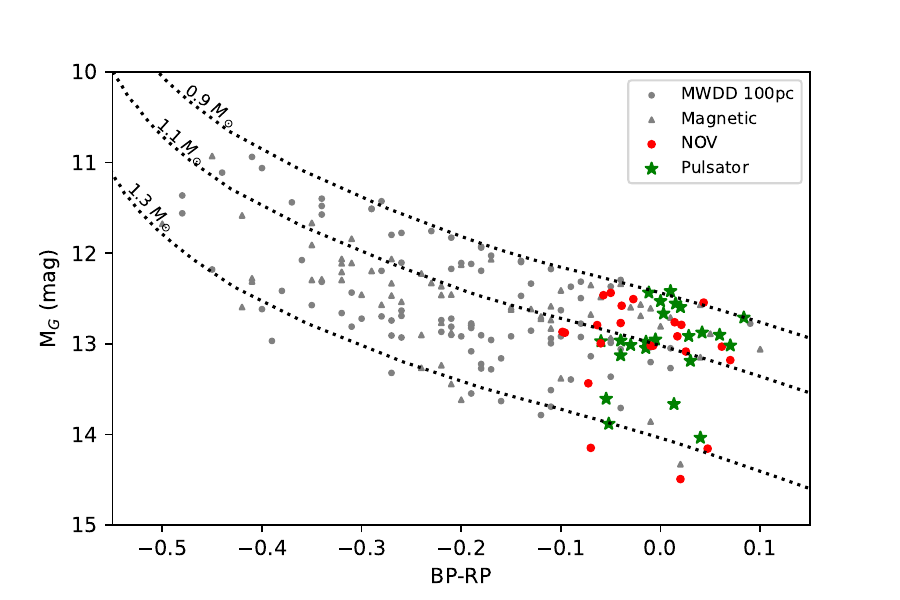}
\caption{Color magnitude diagram of the massive ($M > 0.9~M_\odot$) white dwarfs with $T_{\rm eff}\geq11,000$ K in the MWDD 100 pc sample and the Pan-STARRS footprint from \citet{jewett2024}. The dotted lines show the evolutionary sequences for 0.9, 1.1, and $1.3~M_\odot$ white dwarfs with C/O cores \citep[top to bottom,][]{bergeron19}. The green stars mark both the previously known and the newly discovered pulsating white dwarfs in the sample. Similarly, the red points mark the previously known and newly discovered NOVs.}
%There are magnetic white dwarfs in the MWDD 100 pc sample that fall in or near the instability strip, and those are the gray points that overlap with the pulsators in this diagram.}
\label{fig:hr}
\end{figure}

Figure \ref{fig:hr} shows the Gaia color magnitude diagram with the massive MWDD 100 pc sample in gray, pulsating white dwarfs in green, and NOV white dwarfs in red. The pulsating (green) white dwarf sample includes both newly discovered pulsators presented in this work as well as previously known pulsators in the massive MWDD 100 pc sample. The same is true for the NOV (red) white dwarfs shown. This figure shows
the approximate boundaries of the ZZ Ceti instability strip for massive white dwarfs in the Gaia color-magnitude diagram. The MWDD 100 pc sample
includes magnetic white dwarfs that fall within or near the instability strip, those are the gray triangles that fall near the pulsators in this diagram.
The massive pulsators in our sample are located between $G_{\rm BP}-G_{\rm RP}=-0.06$ to +0.09 and $M_{\rm G}=12$ to 14. 

\subsection{Observations}

Our primary telescope for observations was the 3.5m at the Apache Point Observatory (APO). We used the Astrophysical Research Consortium Telescope Imaging
Camera (ARCTIC) with the BG40 filter and $3\times3$ binning giving us a plate scale of 0.34 arcsec per pixel. We typically obtained 10 s exposures
over 2 hours for most targets. With an overhead of 4.5 s for each exposure, this results in a cadence of 14.5 s. 
The length of observations, significant frequencies, and 4$\langle {\rm A}\rangle$ detection limits where $\langle {\rm A}\rangle$ is the average amplitude
in the Fourier transform for the pulsators and NOVs are presented in Table \ref{tab:freqs} and Table \ref{tab:nov} (see Appendix \ref{appendix:b}), respectively.
 
\begin{deluxetable}{lclrcc}
\tabletypesize{\tiny}
\tablecolumns{6} \tablewidth{0pt}
\caption{Observation details and significant frequencies for the pulsators from APO and NTT.}
\label{tab:freqs}
\tablehead{\colhead{Observation} &  \colhead{Length} & \colhead{Frequency}  & \colhead{Amplitude} &  \colhead{Period} & \colhead{{4$\langle {\rm A}\rangle$}} \\ (UT Date) & (hrs) & (cycles per day) & (mma) & (s) & (mma) }
\startdata 
J0039$-$0357 2024 Oct 9  &2.1 & F1 305.806161 $\pm{0.852963}$ & 8.8 $\pm{1.2}$ & 282.531914 $\pm{0.788046}$ & 6.7 \\
&& F2 318.017123 $\pm{0.232395}$ &  32.3 $\pm{1.2}$ & 271.683484 $\pm{0.198536}$ & \\
&& F3 330.304311 $\pm{0.293023}$ & 25.6 $\pm{1.2}$ & 261.576967 $\pm{0.232053}$ & \\
&& F4 632.896625 $\pm{0.858287}$ & 8.8 $\pm{1.2}$ & 136.515185 $\pm{0.185132}$ & \\
\hline
J0039$-$0357 2024 Nov 6 & 1.8&F1 329.545632 $\pm{0.228252}$ & 33.6 $\pm{1.0}$ & 262.179169 $\pm{0.181592}$ & 4.9\\
&& F2 664.211752 $\pm{1.113189}$ & 6.9 $\pm{1.0}$ &  130.078999 $\pm{0.218007}$ & \\
\hline
J0039$-$0357 2024 Dec 1 &4.0& F1 236.758272 $\pm{0.311809}$ & 7.5 $\pm{0.7}$ & 364.929171 $\pm{0.480609}$& 3.7\\
&& F2 303.799586 $\pm{0.186465}$&  12.6 $\pm{0.7}$& 284.398018 $\pm{0.174557}$ & \\
&& F3 318.330153 $\pm{0.180977}$ &13.0 $\pm{0.7}$ &271.416324 $\pm{0.154306}$& \\
&& F4 327.816169 $\pm{0.190752}$& 12.3 $\pm{0.7}$& 263.562350 $\pm{0.153364}$& \\
&& F5 373.159609 $\pm{0.516541}$& 4.5 $\pm{0.7}$ &231.536313 $\pm{0.320501}$& \\
&& F6 562.340182 $\pm{0.203228}$ &11.5 $\pm{0.7}$& 153.643653 $\pm{0.055526}$ & \\
&& F7 891.439809 $\pm{0.398957}$& 5.9 $\pm{0.7}$ &96.921855 $\pm{0.043377}$ & \\
\hline 
J0039$-$0357 2024 Dec 23 & 3.9&F1 330.929490 $\pm{0.112032}$ & 23.9 $\pm{0.8}$  & 261.082806 $\pm{0.088386}$ & 4.0\\
&& F2 663.787672 $\pm{0.437500}$  & 6.1 $\pm{0.8}$  & 130.162104 $\pm{0.0857894}$ &\\
&& F3 674.621330 $\pm{0.424394}$  & 6.3 $\pm{0.8}$  & 128.071847 $\pm{0.080568}$ &\\ 
\hline
J0154$+$4700 2024 Nov 6 &2.1& F1 190.727239 $\pm{0.336841}$ & 13.8 $\pm{0.7}$ &453.002940 $\pm{0.800043}$&  3.7\\
&& F2 204.947704 $\pm{0.229980}$ &20.2 $\pm{0.7}$ &421.570958 $\pm{0.473062}$ & \\
 \hline
J0158$-$2503 2024 Dec 9 &2.0& F1 185.471956 $\pm{1.267393}$ & 5.4 $\pm{1.1}$ &465.838620 $\pm{3.183234}$ &  5.4 \\
&& F2 205.859362 $\pm{0.229672}$ & 30.1 $\pm{1.1}$ &419.704011 $\pm{0.468253}$ & \\
\hline 
J0158$-$2503 2024 Dec 25 &3.5& F1 195.827847 $\pm{0.216539}$ &11.5 $\pm{0.7}$ &441.203850 $\pm{0.487866}$& 3.2\\
&& F2 205.333074 $\pm{0.054050}$ &46.2 $\pm{0.7}$& 420.779752 $\pm{0.110762}$ & \\
&&F3 216.696535 $\pm{0.119549}$ &20.9 $\pm{0.7}$ &398.714266 $\pm{0.219966}$ & \\
&&F4 411.510990 $\pm{0.267276}$&9.3 $\pm{0.7}$ &209.957941 $\pm{0.136367}$  & \\
&&F5 422.210353 $\pm{0.349874}$ &7.1 $\pm{0.7}$ &204.637332 $\pm{0.169577}$  & \\
\hline 
J0204$+$8713 2023 Apr 15 &2.1& F1 245.025464 $\pm{1.166706}$ & 5.0 $\pm{0.9}$ &352.616412 $\pm{1.679008}$& 4.5\\
&& F2 253.522357 $\pm{1.170946}$&  5.0 $\pm{0.9}$ & 340.798346 $\pm{1.574048}$ & \\
\hline 
J0204$+$8713 2023 Nov 24 &3.5& F1 219.381927 $\pm{0.673149}$ & 5.5 $\pm{1.0}$ & 393.833718 $\pm{1.208435}$& 5.0\\
&& F2 227.993568 $\pm{0.634822}$ & 5.8 $\pm{1.0}$ & 378.958059 $\pm{1.055165}$ & \\
&& F3 260.311874 $\pm{0.695208}$ & 5.3 $\pm{1.0}$ & 331.909562 $\pm{0.886422}$ & \\
\hline 
J0204$+$8713 2023 Dec 30 &3.3& F1 266.602480 $\pm{0.669197}$ & 6.1 $\pm{1.0}$ & 324.078006 $\pm{0.813466}$& 5.2\\
\hline 
J0204$+$8713 2024 Mar 10 &2.8& F1 200.571783 $\pm{1.031661}$ & 3.5 $\pm{0.8}$ & 430.768470 $\pm{2.215701}$& 3.8\\
&& F2 231.880286 $\pm{0.560326}$ & 6.4 $\pm{0.8}$ & 372.606061 $\pm{0.900382}$ & \\
&& F3 240.743365 $\pm{0.427515}$ & 8.4 $\pm{0.8}$ & 358.888396 $\pm{0.637318}$ & \\
&& F4 245.832988 $\pm{0.396233}$ & 9.1 $\pm{0.8}$ & 351.458121 $\pm{0.566479}$ & \\
\hline 
J0204$+$8713 2024 Apr 20 &2.8& F1 257.403522 $\pm{0.753537}$ & 7.0 $\pm{1.1}$ & 335.659743 $\pm{0.982628}$& 5.7\\
\hline 
J0634$+$3848 2024 Dec 9 &2.0& F1 839.842921 $\pm{1.147355}$ & 1.5 $\pm{0.3}$ & 102.876381 $\pm{0.140545}$ & 1.3\\
\hline 
J0634$+$3848 2024 Dec 23 &1.8& F1 41.138520 $\pm{0.215606}$ & 7.9 $\pm{0.2}$ & 2100.221398 $\pm{11.007210}$ & 1.1\\
&& F2 488.564003 $\pm{1.078713}$ & 1.6 $\pm{0.2}$ & 176.844793 $\pm{0.390460}$ & \\
&& F3 838.273053 $\pm{1.143435}$ & 1.5 $\pm{0.2}$ & 103.069041 $\pm{0.140590}$ & \\
\hline
J0634$+$3848 2025 Feb 3 &2.7& F1 39.988517 $\pm{0.314764}$ & 2.8 $\pm{0.2}$ & 2160.620260 $\pm{17.007019}$ & 0.9\\
&& F2 489.604041 $\pm{0.838079}$ & 1.1 $\pm{0.2}$ & 176.469132 $\pm{0.302071}$ & \\ 
&& F3 839.930123 $\pm{0.687973}$ & 1.3 $\pm{0.2}$ & 102.865700 $\pm{0.084256}$ & \\
\hline 
J0712$-$1815 2025 Jan 19 &1.7&  F1 83.947331 $\pm{1.019242}$ & 7.1 $\pm{0.9}$ & 1029.216760 $\pm{12.496180}$ & 4.3\\
&& F2 151.708404 $\pm{0.988647}$ & 7.3 $\pm{0.9}$ & 569.513605 $\pm{3.711383}$ & \\
&& F3 173.764146 $\pm{0.444767}$ & 16.2 $\pm{0.9}$ & 497.225705 $\pm{1.272700}$ & \\
&& F4 205.496959 $\pm{0.582887}$ & 12.4 $\pm{0.9}$ & 420.444178 $\pm{1.192579}$ & \\
&& F5 256.850678 $\pm{0.451733}$ & 16.0 $\pm{0.9}$ & 336.382215 $\pm{0.591608}$ & \\
\hline 
J0712$-$1815 2025 Feb 5 &2.1&  F1 85.372940 $\pm{0.888821}$ & 5.0 $\pm{0.7}$ &1012.030276 $\pm{10.536287}$ & 3.2\\
&& F2 152.917895 $\pm{0.786854}$ &5.6 $\pm{0.7}$ &565.009085 $\pm{2.907310}$ & \\
&& F3 170.661015 $\pm{0.371232}$ &12.0 $\pm{0.7}$ &506.266765 $\pm{1.101262}$  & \\
&& F4 205.650495 $\pm{0.317435}$ &14.0 $\pm{0.7}$ &420.130280 $\pm{0.648499}$ & \\
&& F5 255.632404 $\pm{0.247233}$ &18.0 $\pm{0.7}$ &337.985321 $\pm{0.326880}$ & \\
\hline 
J0912$-$2642 2024 Nov 23 &2.5&  F1 85.319740 $\pm{0.705133}$ & 2.3 $\pm{0.3}$ &1012.661314 $\pm{8.369234}$ & 1.5\\
&& F2 91.748870 $\pm{0.884533}$ &1.8 $\pm{0.3}$ &941.700971 $\pm{9.078756}$ & \\
\hline
J0959$-$1828 2025 Feb 20 &1.9&  F1 401.207772 $\pm{0.785705}$ & 7.4 $\pm{0.8}$ & 215.349766 $\pm{0.421730}$ & 4.3\\
\hline 
J0959$-$1828 2025 Mar 28 &4.1&  F1 402.353479 $\pm{0.406530}$ & 0.7 $\pm{0.1}$ & 214.736555 $\pm{0.216966}$ & 0.4\\
\hline
J1052$+$1610 2025 Feb 20 &2.1&  F1 110.100033 $\pm{0.703626}$ & 5.0 $\pm{0.5}$ & 784.740909 $\pm{5.015113}$ & 2.7\\
&& F2 125.465008 $\pm{0.736707}$ & 4.8 $\pm{0.5}$ & 688.638222 $\pm{4.043555}$ & \\
\hline 
J1451$-$2502 2025 Feb 20 &1.3&  F1 152.394438 $\pm{1.314322}$ & 9.4 $\pm{1.2}$ & 566.949825 $\pm{4.889645}$ & 6.1\\
&& F2 181.889447 $\pm{0.515562}$ & 23.9 $\pm{1.2}$ & 475.013814 $\pm{1.346417}$ & \\
\hline 
J1626$+$2533 2024 May 18 &2.2&  F1 59.772653 $\pm{1.110704}$ & 3.2 $\pm{0.6}$ & 1445.477081 $\pm{26.860062}$ & 2.9 \\
\hline 
J1626$+$2533 2024 Aug 30 &1.8&  F1 55.157900 $\pm{0.865205}$ & 3.6 $\pm{0.4}$ & 1566.412064 $\pm{24.570688}$ & 2.2 \\
\hline 
J1722$+$3958 2024 Aug 8&&  F1 180.570526 $\pm{0.856759}$ &8.1 $\pm{1.0}$ &478.483404 $\pm{2.270276}$ & 5.5\\
&& F2 221.459554 $\pm{0.603540}$ &11.4 $\pm{1.0}$ &390.138960 $\pm{1.063239}$ & \\
&& F3 241.467530 $\pm{0.350202}$ &19.7 $\pm{1.0}$ &357.812083 $\pm{0.518937}$ & \\
\hline
J1929$-$2926 2024 Aug 8 &2.1& F1 201.690586 $\pm{0.613081}$ & 10.0 $\pm{1.0}$ & 428.378943 $\pm{1.302148}$ & 4.7\\
&& F2 231.051521 $\pm{0.860352}$ & 7.2 $\pm{1.0}$ & 373.942572 $\pm{1.392426}$ & \\
\hline 
J2026$-$2254 2024 Jun 14 &1.4&  F1 173.971069 $\pm{0.545665}$ & 35.8 $\pm{2.1}$ & 496.634300 $\pm{1.557707}$ & 10.3\\
&& F2 272.089417 $\pm{0.698970}$ & 11.5 $\pm{2.1}$ & 317.542670 $\pm{1.982787}$ & \\
\hline
J2208$+$2059 2024 Aug 30 &2.0&  F1 139.778117 $\pm{0.484904}$ &5.2 $\pm{0.4}$ &618.122506 $\pm{2.144328}$ & 1.9\\
&& F2 157.468572 $\pm{0.343530}$ &7.4 $\pm{0.4}$ &548.680914 $\pm{1.196990}$ & \\
&& F3 182.053632 $\pm{1.154111}$ &2.2 $\pm{0.4}$ &474.585423 $\pm{3.008587}$ & \\
&& F4 200.057685 $\pm{0.658022}$ &3.9 $\pm{0.4}$ &431.875436 $\pm{1.420508}$ & \\
\hline 
\enddata    
\end{deluxetable}

\begin{deluxetable}{lclrcc}
\tabletypesize{\tiny}
\tablecolumns{6} \tablewidth{0pt}
\caption{Observation details and significant frequencies for the pulsators from GTC.}
\label{tab:freqs_hiper}
\tablehead{\colhead{Observation} &  \colhead{Length} & \colhead{Frequency}  & \colhead{Amplitude} &  \colhead{Period} & \colhead{{4$\langle {\rm A}\rangle$}} \\ (UT Date \& Band) & (hrs) & (cycles per day) & (mma) & (s) & (mma) }
\startdata 
J0039$-$0357 2024 Nov 11 & 4.0 & F1 319.445760 $\pm{0.354900}$ & 4.1 $\pm{0.4}$ & 270.468451 $\pm{0.300487}$ & 2.3\\
$u$-band & & F2 331.132800 $\pm{0.092425}$ & 15.6 $\pm{0.4}$ & 260.922506 $\pm{0.072828}$ & \\
& & F3 657.770584 $\pm{0.528995}$ & 2.7 $\pm{0.4}$ & 131.352788 $\pm{0.105334}$ & \\
\hline 
J0039$-$0357 2024 Nov 11 & 4.0 & F1  319.221015 $\pm{0.343340}$ & 4.6 $\pm{0.5}$ & 270.658873 $\pm{0.327714}$ & 2.7\\
$g$-band & & F2 330.899833 $\pm{0.076044}$ & 20.7 $\pm{0.5}$ & 261.106206 $\pm{0.060005}$ & \\
& & F3 658.206184 $\pm{0.402211}$ & 3.9 $\pm{0.5}$ & 131.265859 $\pm{0.080213}$ & \\
\hline 
J0039$-$0357 2024 Nov 11 & 4.0 & F1 319.366661 $\pm{0.394019}$ & 2.0 $\pm{0.2}$ & 270.535440 $\pm{0.333773}$ & 1.2\\
$r$-band& & F2 331.321595 $\pm{0.122158}$ & 6.6 $\pm{0.2}$ & 260.773826 $\pm{0.096147}$ & \\
\hline 
J0039$-$0357 2024 Nov 11 & 4.0 & F1 320.355050 $\pm{0.576934}$ & 1.1 $\pm{0.2}$ & 269.700759 $\pm{0.485710}$ &  1.0\\
$i$-band& & F2 330.937644 $\pm{0.213182}$ & 3.0 $\pm{0.2}$ & 261.076374 $\pm{0.168179}$ & \\
\hline 
J0039$-$0357 2024 Nov 11 & 4.0 & F1 331.538544 $\pm{0.347016}$ & 1.9 $\pm{0.2}$ & 260.603183 $\pm{0.272769}$ &  1.0\\
$z$-band & & & & & \\
\hline
J0204+8713 2024 Oct 24 & 2.8 &  F1 229.049997 $\pm{0.549858}$ & 2.4 $\pm{0.3}$ & 377.210221 $\pm{0.905532}$ & 1.4 \\
$u$-band & & F2 241.229508 $\pm{0.401153}$ & 3.3 $\pm{0.3}$ & 358.165138 $\pm{0.595611}$ & \\
& & F3 263.272892 $\pm{0.527690}$ & 2.5 $\pm{0.3}$ & 328.176590 $\pm{0.657779}$ & \\
\hline 
J0204+8713 2024 Oct 24 & 2.8 & F1 210.070146 $\pm{0.367653}$ & 3.4 $\pm{0.3}$ & 411.291188 $\pm{0.719819}$ & 1.9 \\
$g$-band & & F2 228.423821 $\pm{0.343547}$ & 3.6 $\pm{0.3}$ & 378.244264 $\pm{0.568875}$ & \\
& & F3 240.453547 $\pm{0.209675}$ & 5.9 $\pm{0.3}$ & 359.320963 $\pm{0.313327}$ & \\
& & F4 261.344304 $\pm{0.309773}$ & 4.0 $\pm{0.3}$ & 330.598367 $\pm{0.391860}$ & \\
\hline 
J0204+8713 2024 Oct 24 & 2.8 & F1 209.669235 $\pm{0.459851}$ & 4.0 $\pm{0.4}$ & 412.077623 $\pm{0.903777}$ & 2.2 \\
$r$-band & & F2 241.269079 $\pm{0.305188}$ & 5.5 $\pm{0.4}$ & 358.106395 $\pm{0.452979}$ & \\
& & F3 227.177256 $\pm{0.533518}$ & 3.1 $\pm{0.4}$ & 380.319762 $\pm{.893168}$ & \\
& & F4 261.766275 $\pm{0.555513}$ & 2.9 $\pm{0.4}$ & 330.065437 $\pm{0.700456}$ & \\
\hline 
J0204+8713 2024 Oct 24 & 2.8 & F1 209.242210 $\pm{0.542543}$ & 1.3 $\pm{0.2}$ & 412.918598 $\pm{1.070654}$ & 0.9 \\
$i$-band & & F2 227.604281 $\pm{0.759877}$ & 1.0 $\pm{0.2}$ & 379.606217 $\pm{1.267349}$ & \\
& & F3 241.269079 $\pm{0.348883}$ & 2.1 $\pm{0.2}$ & 358.106395 $\pm{0.517834}$ & \\
& & F4 263.474374 $\pm{0.502134}$ & 1.4 $\pm{0.2}$ & 327.92563 $\pm{0.624966}$ & \\
\hline 
J0204+8713 2024 Oct 24 & 2.8 & F1 182.377506 $\pm{0.774661}$ & 0.7 $\pm{0.1}$ & 473.742634 $\pm{2.012254}$ & 0.67 \\
$z$-band & & F2 225.131904 $\pm{0.832872}$ & 0.7 $\pm{0.1}$ & 383.775016 $\pm{1.41977}$ & \\
& & F3 241.532812 $\pm{0.623171}$ & 0.9 $\pm{0.1}$ & 357.715373 $\pm{0.922930}$ & \\
\hline 
J0959$-$1828 2025 Jan 28 & 2.0 & F1 405.322396 $\pm{0.392189}$ & 5.5 $\pm{0.3}$ & 213.163647 $\pm{0.206257}$ & 1.7 \\
$u$-band & & F2 431.073071 $\pm{0.920288}$ & 2.3 $\pm{0.3}$ & 200.430057 $\pm{0.427894}$ & \\
\hline 
J0959$-$1828 2025 Jan 28 & 2.0 & F1 404.458573 $\pm{0.170671}$ & 1.3 $\pm{0.03}$ & 213.618911 $\pm{0.090142}$ & 0.2 \\
$g$-band & & F2 430.181015 $\pm{0.437800}$ & 0.5 $\pm{0.03}$ & 200.845684 $\pm{0.204403}$ & \\
\hline 
J0959$-$1828 2025 Jan 28 & 2.0 & F1  403.432286 $\pm{0.270891}$ & 2.0 $\pm{0.1}$ & 214.162334 $\pm{0.143803}$ & 0.5 \\
$r$-band & & F2  442.490778 $\pm{0.709953}$ & 0.8 $\pm{0.1}$ & 195.258307 $\pm{0.313282}$ & 0.5\\
\hline 
J0959$-$1828 2025 Jan 28 & 2.0 & F1 403.763705 $\pm{0.406464}$ & 0.9 $\pm{0.1}$ & 213.986544 $\pm{0.215418}$ & 0.3 \\
$i$-band & & F2 430.096437 $\pm{0.965571}$ & 0.4 $\pm{0.1}$ & 200.88518 $\pm{0.450989}$ & \\
\hline 
J0959$-$1828 2025 Jan 28 & 2.0 & F1 402.470662 $\pm{0.984905}$ & 0.5 $\pm{0.1}$ & 214.674033 $\pm{0.525339}$ & 0.4 \\
$z$-band & & & & & \\ 
\hline 
J1106$+$1802 2025 Jan 12 & 4.0 & F1 159.057435 $\pm{0.449416}$ & 9.5 $\pm{1.3}$ & 543.200008 $\pm{1.534809}$ & 6.2 \\
$u$-band & & F2 233.947351 $\pm{0.043353}$ & 98.7 $\pm{1.3}$ & 369.313863 $\pm{0.068438}$ & \\
& & F3 468.006138 $\pm{0.108386}$ & 39.4 $\pm{1.3}$ & 184.612963 $\pm{0.042755}$ & \\
& & F4 701.940093 $\pm{0.278442}$ & 15.3 $\pm{1.3}$ & 123.087427 $\pm{0.048826}$ & \\
\hline 
J1106$+$1802 2025 Jan 12 & 4.0 & F1 159.218326 $\pm{0.190364}$ & 5.2 $\pm{0.3}$ & 542.651102 $\pm{0.648802}$ & 2.0 \\
$g$-band & & F2 233.950087 $\pm{0.020187}$ & 48.9 $\pm{0.3}$ &  369.309544 $\pm{0.031867}$ & \\
& & F3 467.959894 $\pm{0.051295}$ & 19.2 $\pm{0.3}$ & 184.631207 $\pm{0.020238}$ & \\
& & F4 702.105212 $\pm{0.123871}$ & 8.0 $\pm{0.3}$ & 123.05848 $\pm{0.021711}$ & \\
\hline 
J1106$+$1802 2025 Jan 12 & 4.0 & F1 159.133772 $\pm{0.248271}$ & 2.7 $\pm{0.2}$ & 542.939433 $\pm{0.847062}$ & 1.0 \\
$r$-band & & F2 233.964876 $\pm{0.026701}$ & 24.7 $\pm{0.2}$ & 369.286200 $\pm{0.042144}$ & \\
& & F3 467.976975 $\pm{0.070249}$ & 9.4 $\pm{0.2}$ & 184.624468 $\pm{0.027714}$ & \\
& & F4 702.164688 $\pm{0.178092}$ & 3.7 $\pm{0.2}$ & 123.048056 $\pm{0.031209}$ & \\
\hline 
J1106$+$1802 2025 Jan 12 & 4.0 & F1 159.319339 $\pm{0.299202}$ & 1.6 $\pm{0.1}$ & 542.307045 $\pm{1.018454}$ & 0.8  \\
$i$-band & & F2 233.902946 $\pm{0.031397}$ & 15.5 $\pm{0.1}$ & 369.383975 $\pm{0.049583}$ & \\
& & F3 468.043623 $\pm{0.084273}$ & 5.8 $\pm{0.1}$ & 184.598178 $\pm{0.033238}$ & \\ 
& & F4 702.145067 $\pm{0.207335}$ & 2.3 $\pm{0.1}$ & 123.051495 $\pm{0.036336}$ & \\
\hline 
J1106$+$1802 2025 Jan 12 & 4.0 & F1 159.347339 $\pm{0.496017}$ & 1.0 $\pm{0.2}$ & 542.211753 $\pm{1.687799}$ & 0.8  \\
$z$-band & & F2 233.912839 $\pm{0.049589}$ & 10.4 $\pm{0.2}$ & 369.368353 $\pm{0.078305}$ & \\
& & F3 467.785212 $\pm{0.126927}$ & 4.1 $\pm{0.2}$ & 184.700153 $\pm{0.050116}$ & \\
& & F4 701.797556 $\pm{0.316585}$ & 1.6 $\pm{0.2}$ & 123.112426 $\pm{0.055537}$ & \\
\hline 
\enddata    
\end{deluxetable}

\addtocounter{table}{-1}
\begin{deluxetable}{lclrcc}
\tabletypesize{\tiny}
\tablecolumns{6} \tablewidth{0pt}
\caption{Continued.}
\label{tab:freqs_hiper}
\tablehead{\colhead{Observation} &  \colhead{Length} & \colhead{Frequency}  & \colhead{Amplitude} &  \colhead{Period} & \colhead{{4$\langle {\rm A}\rangle$}} \\ (UT Date \& Band) & (hrs) & (cycles per day) & (mma) & (s) & (mma) }
\startdata 
J1106$+$1802 2025 Jan 19 & 4.0 & F1 161.107075 $\pm{0.373447}$ & 14.2 $\pm{1.6}$ & 536.289297  $\pm{1.243121}$ & 6.8 \\
$u$-band & & F2 173.408466 $\pm{0.417985}$ & 12.8 $\pm{1.6}$ & 498.24557 $\pm{1.200975}$ & \\
& & F3 234.170136 $\pm{0.050900}$ & 101.2 $\pm{1.5}$ & 368.962505 $\pm{0.080199}$ & \\
& & F4 294.806456 $\pm{0.364521}$ & 14.1 $\pm{1.5}$ & 293.073636 $\pm{0.362378}$ & \\
& & F5 468.012829 $\pm{0.133153}$ & 38.4 $\pm{1.5}$ & 184.610324 $\pm{0.052523}$ & \\
& & F6 702.047175 $\pm{0.331008}$ & 15.4 $\pm{1.5}$ & 123.068653 $\pm{0.058026}$ & \\
\hline 
J1106$+$1802 2025 Jan 19 & 4.0 & F1 161.320482 $\pm{0.164128}$ & 6.8 $\pm{0.3}$ & 535.579853 $\pm{0.544901}$ & 2.1 \\
$g$-band & & F2 173.406241 $\pm{0.175441}$ & 6.3 $\pm{0.3}$ & 498.251963 $\pm{0.504098}$ & \\
& & F3 234.157626 $\pm{0.022331}$ & 49.6 $\pm{0.3}$ & 368.982217 $\pm{0.035189}$ & \\
& & F4 295.041665 $\pm{0.162018}$ & 6.8 $\pm{0.3}$ & 292.839996 $\pm{0.160809}$ & \\
& & F5 467.985148 $\pm{0.060464}$ & 18.3 $\pm{0.3}$ & 184.621244 $\pm{0.023853}$ & \\
& & F6 702.052278 $\pm{0.146153}$ & 7.6 $\pm{0.3}$ & 123.067758 $\pm{0.025620}$ & \\
\hline 
J1106$+$1802 2025 Jan 19 & 4.0 & F1 161.201335 $\pm{0.226324}$ & 3.4 $\pm{0.2}$ & 535.975710 $\pm{0.752501}$ & 1.1 \\
$r$-band & & F2 173.409968 $\pm{0.246289}$ & 3.2 $\pm{0.2}$ & 498.241255 $\pm{0.707637}$ & \\
& & F3 234.171916 $\pm{0.031167}$ & 25.0 $\pm{0.2}$ & 368.959701 $\pm{0.049107}$ & \\
& & F4 294.944749 $\pm{0.220718}$ & 3.5 $\pm{0.2}$ & 292.936220 $\pm{0.219215}$ & \\
& & F5 468.017662 $\pm{0.087305}$ & 8.9 $\pm{0.2}$ & 184.608418 $\pm{0.034437}$ & \\
& & F6 702.196743 $\pm{0.222142}$ & 3.5 $\pm{0.2}$ & 123.042439 $\pm{0.038925}$ & \\
\hline 
J1106$+$1802 2025 Jan 19 & 4.0 & F1 161.194328 $\pm{0.221579}$ & 2.4 $\pm{0.2}$ & 535.999009 $\pm{0.736788}$ & 0.8 \\
$i$-band & & F2 173.378494 $\pm{0.277647}$ & 1.9 $\pm{0.2}$ & 498.331702 $\pm{0.798025}$ & \\
& & F3 234.182128 $\pm{0.034076}$ & 15.6 $\pm{0.2}$ & 368.943611 $\pm{0.053685}$ & \\
& & F4 295.044837 $\pm{0.251900}$ & 2.1 $\pm{0.2}$ & 292.836848 $\pm{0.250015}$ & \\
& & F5 467.974379 $\pm{0.097340}$ & 5.4 $\pm{0.2}$ & 184.625492 $\pm{0.038403}$ & \\
& & F6 702.215787 $\pm{0.237386}$ & 2.2 $\pm{0.2}$ & 123.039102 $\pm{0.041594}$ & \\
\hline 
J1106$+$1802 2025 Jan 19 & 4.0 & F1 234.158865 $\pm{0.058860}$ & 10.5 $\pm{0.2}$ & 368.980265 $\pm{0.092750}$ & 0.9 \\
$z$-band & & F2 295.189508 $\pm{0.438546}$ & 1.4 $\pm{0.2}$ & 292.693330 $\pm{0.434838}$ & \\
& & F3 468.050567 $\pm{0.165373}$ & 3.8 $\pm{0.2}$ & 184.595439 $\pm{0.065222}$ & \\
& & F4 702.236774 $\pm{0.392176}$ & 1.6 $\pm{0.2}$ & 123.035425 $\pm{0.068711}$ & \\
\hline 
\enddata    
\end{deluxetable}

For one of our targets, J0959$-$1828, we were able to obtain follow up observations with ULTRACAM \citep{dhillon}, an ultra fast CCD camera, on the
3.5m ESO New Technology Telescope (NTT) on UT 2025 Mar 28. ULTRACAM uses a triple beam setup and three frame transfer CCD cameras, which allows
simultaneous data in three different filters with negligible (24 ms) dead-time between exposures. We used the super-SDSS $u_s$, $g_s$, and $r_s$ filters \citep{brown}.
ULTRACAM observations were taken over 4 hours. 

We were able to observe 6 of our targets (J0039$-$0357, J0204+8713, J0408$+$2323, J0657$+$7341, J0959$-$1828, and J1106$+$1802) using HiPERCAM \citep{dhillon_hipercam}, a high speed quintuple-beam CCD camera, on the 10.4 m Gran Telescopio Canarias (GTC).  We used the super-SDSS $u_s$, $g_s$, $r_s$, $i_s$, and $z_s$ filters. All of the observations taken were over 4 hours, with the exceptions of J0408$+$2323 and J0959$-$1828.

\subsection{Analysis}

We used the {\tt Period4} software \citep{lenz} to generate a Fourier transform for each night of observation in order to determine the significant pulsation frequencies. We applied a non-linear least squares fitting procedure, where we identify the most significant peak
above the detection threshold, fit it, and subtract the corresponding sinusoidal signal from the data, and repeat the process until no additional peaks
exceeding the 4$\langle {\rm A}\rangle$ detection threshold are found. This detection threshold is defined as 4 times the average amplitude of the Fourier transform, which serves as the minimum amplitude for a frequency to be
considered significant \citep{breger, degeronimo, 2025ApJ...988...32C}.This allowed us to accurately determine the frequency and amplitude of each
pulsation mode. For observations with multiple bands, we process all of the data individually and include the frequencies detected in each filter
in Table \ref{tab:freqs_hiper}. We discuss the pulsating white dwarf sample in the next section, and then present white dwarfs not-observed-to-vary in the following
section.

\section{Pulsating White Dwarfs}

\begin{figure*}
\centering
\vspace{-0.2in}
\includegraphics[width=0.9\linewidth]{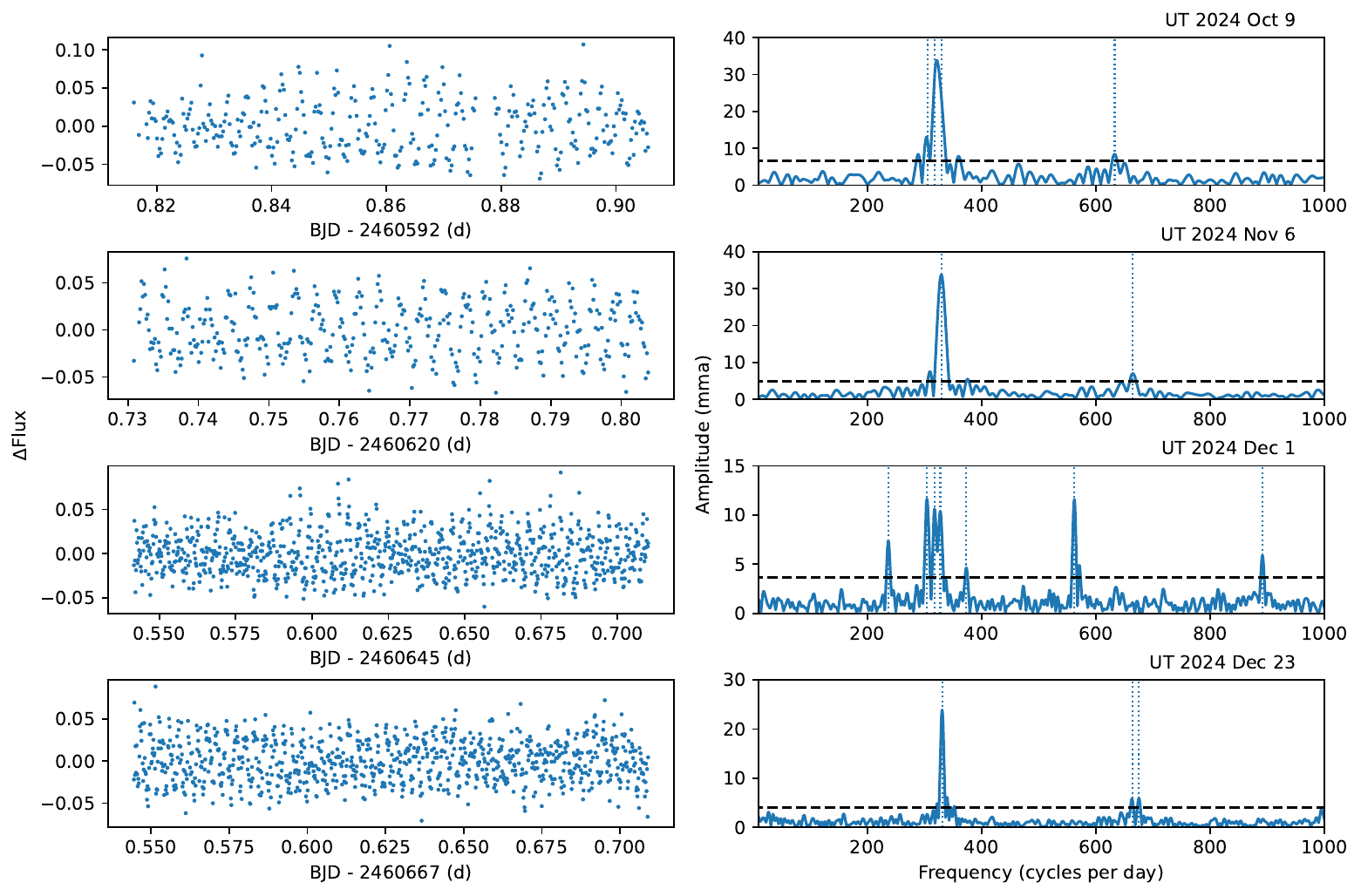}
\caption{Light curves (left) and their Fourier transforms (right panels) for J0039$-$0357 observed on 4 different nights. The horizontal black dashed
lines indicate 4 times the average amplitude in the Fourier transforms, while the vertical dashed blue lines mark the significant frequencies.
Table \ref{tab:freqs} provides a list of the frequencies detected for each observation.}
\label{fig:j0039}
\end{figure*}

\begin{figure*}
\centering
\vspace{-0.2in}
\includegraphics[width=0.9\linewidth]{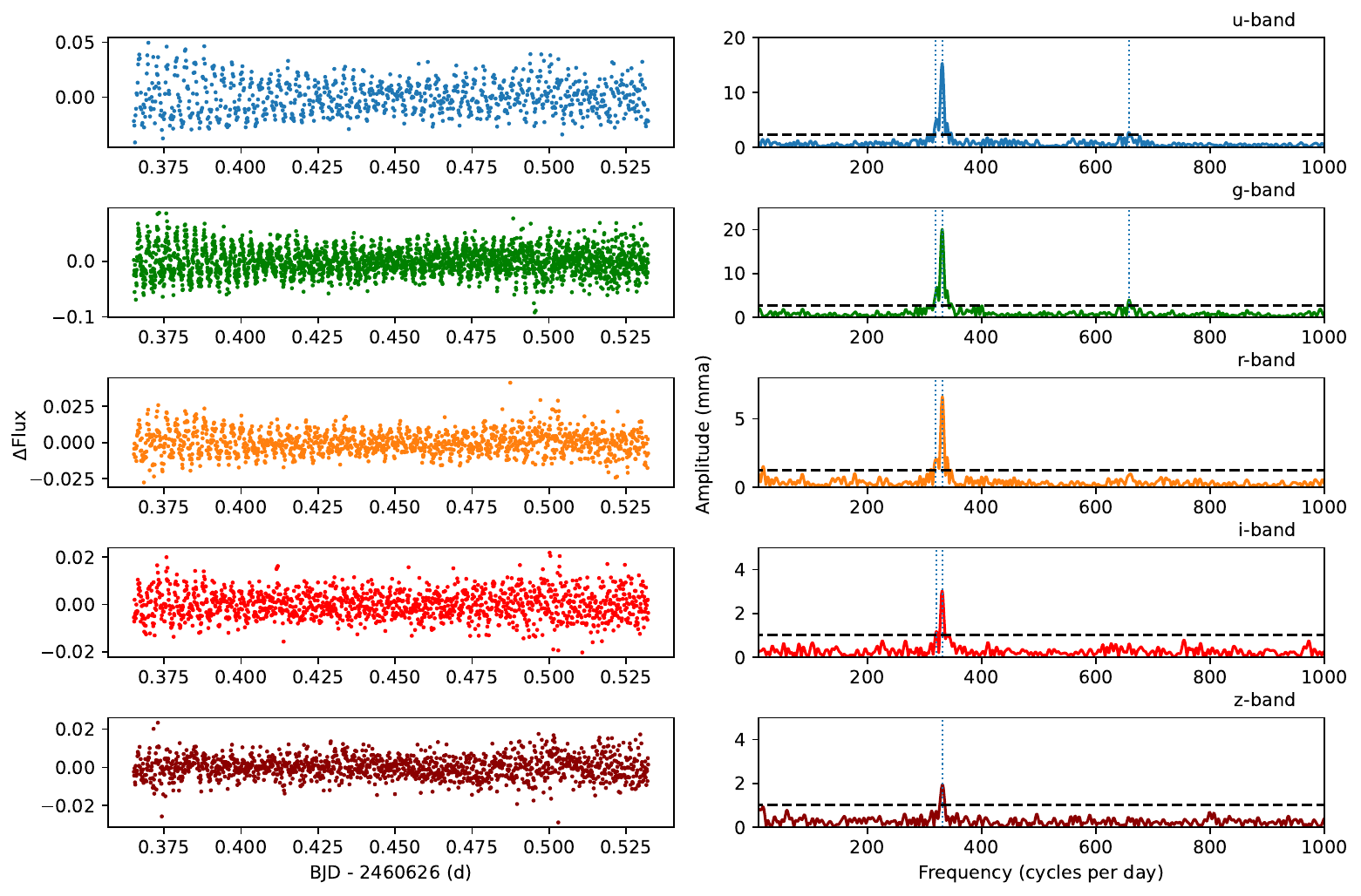}
\caption{Light curves (left) and their Fourier transforms (right panels) for J0039$-$0357 observed on 2024 Nov 11 at the GTC using HiPERCAM. The corresponding filter is labeled above the Fourier Transform. Table \ref{tab:freqs_hiper} provides a list of the frequencies detected for each observation.}
\label{fig:j0039_hiper}
\end{figure*}

With our photometric followup,
we confirm pulsations in 14 new massive white dwarfs. 
\citet{jewett2024} sample contains 10 previously known massive pulsating ZZ Ceti white dwarfs, including the most massive
pulsator previously known J0049$-$2525 \citep{kilic23b,2025ApJ...988...32C} and the recently presented J0135+5722 \citep{degeronimo}. 
The frequencies we deem significant (amplitudes above the 4$\langle {\rm A}\rangle$ detection threshold) are marked by dashed blue lines on the Fourier transforms presented below and are
listed in Table \ref{tab:freqs} along with their amplitudes and periods.  The masses and effective temperatures for each white dwarf are adopted from
\citet{jewett2024} and can be found in Table \ref{tab:obsphys}. 

In this section, we present the light curves and corresponding Fourier transforms for six systems with $M>1.1~M_\odot$. The remaining systems are
presented in Appendix \ref{appendix:a}. 

\begin{figure*}
\centering
\vspace{-0.2in}
\includegraphics[width=0.9\linewidth]{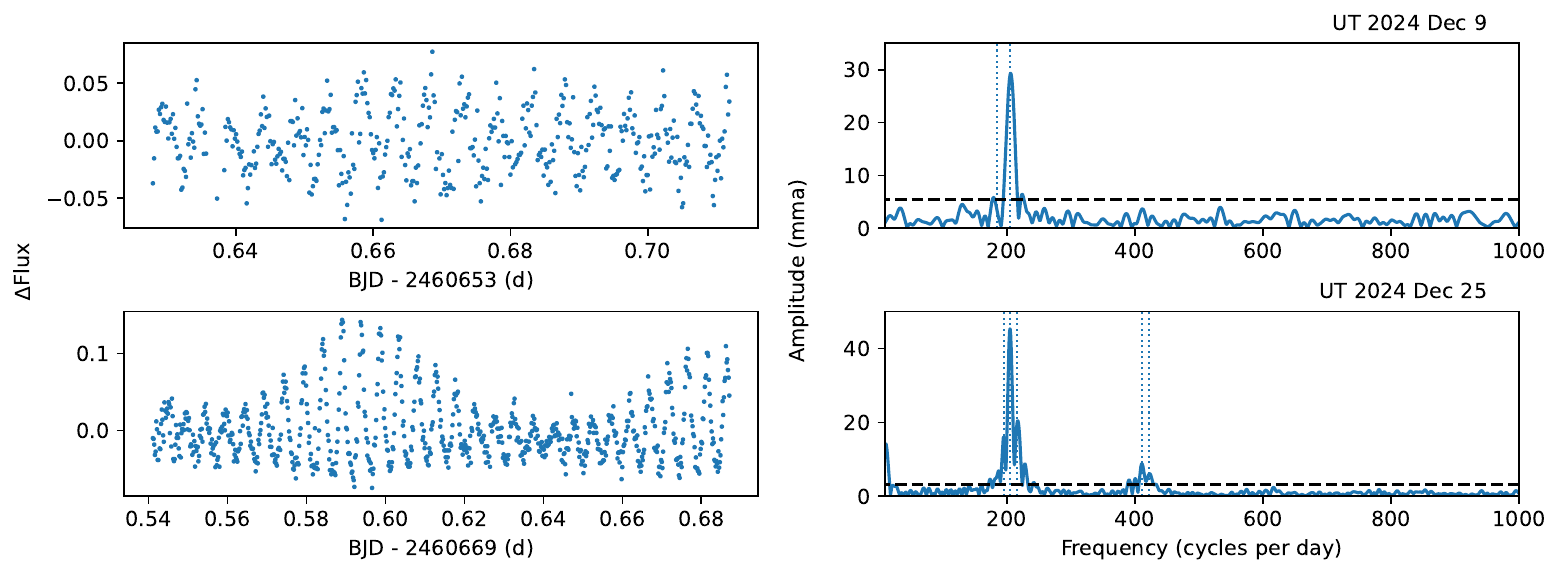}
\caption{Light curves and Fourier transforms for J0158$-$2503.}
\label{fig:j0158}
\end{figure*}

\subsection{J0039$-$0357}

J0039$-$0357 has $M=1.271 \pm 0.009~M_\odot$ and $T_{\rm eff} = 11,871 \pm 214$ K. It is one of the most interesting pulsators that we found as we see
changing pulsation modes on four different nights of observations. The light curves and corresponding Fourier transforms are shown in
Figure \ref{fig:j0039}. On the first night of observation, UT 2024 Oct 9, we detect 4 modes with frequencies of 305.8, 318.0, 330.3, and 632.9 c/d. The mode at
632.9 c/d is likely a combination frequency of the modes at 305.8 and 330.3 c/d as the mode is the sum of these two frequencies. The highest amplitude we detect is for the mode at 318.0 c/d at 32.3 milli-modulation amplitude (mma). The next night of observations on UT 2024 Nov 6, this drops down to just 2 modes detected with frequencies of 329.5 and
664.2 c/d, with the latter being a harmonic ($2\times$) of the former. The first mode has an amplitude of 33.6 mma. 

The third night of observations on UT 2024 Dec 1 shows the richest pulsation spectrum for this object. We detect 7 modes with frequencies ranging
from 236.8 to 891.4 c/d. Several of these modes are likely combination frequencies. For example, the mode at 562.3 c/d is a combination of
236.8 and 327.8 c/d, and 891.4 c/d is a combination of 327.8 and 562.3 c/d. The highest amplitude mode is at 318.3 c/d with an amplitude of 13.0 mma. 
Then, on UT 2024 Dec 23, we find 3 modes with frequencies of 330.9, 663.8, and 674.6 c/d, with the last two likely being combination frequencies. The first mode at 330.9 c/d has the highest amplitude of 23.9 mma. Finally, we also observed J0039$-$0357 at GTC shown in Figure \ref{fig:j0039_hiper}. The $g$-band data shows modes at 319.2, 330.9, and 658.2 c/d. The dominant mode at 330.9 c/d is detected in each filter, with decreasing amplitudes in the redder filters, and
its harmonic at 658.2 c/d is detected in the $u-$ and $g-$bands. 
Clearly, we are seeing some modes disappear and some appear on different nights. The mode at $\approx$ 330 c/d is present in each night of observation,
but with amplitudes ranging from 12 to 34 mma. This seems to be the only mode that we can consistently detect during each observation.

\subsection{J0158$-$2503}

J0158$-$2503 has $M=1.122 \pm 0.007~M_\odot$ and $T_{\rm eff} = 12,234 \pm 94$ K. We observed this white dwarf on two different nights, UT 2024 Dec 9
and 25. The light curves and corresponding Fourier transforms are shown in Figure \ref{fig:j0158}. On the first night of observing, we find
there to be two significant modes with frequencies of 185.5 and 205.9 c/d. The amplitudes of these modes are 5.4 and 30.1 mma, respectively. 

Interestingly, on the second night of observations, we detect 5 modes with frequencies of 195.8, 205.3, 216.7, 411.5, and 422.2 c/d. The dominant mode
at 205.3 c/d has an amplitude of 46.2 mma. Two of the modes are likely combination frequencies: 411.5 c/d is a combination of 195.8 and 205.3 c/d, and 422.2 c/d is a 
combination of 205.3 and 216.7 c/d. We see the mode at $\approx$ 205 c/d  on both nights of observations, but the amplitudes clearly change on each night.  

\begin{figure*}
\centering
\vspace{-0.2in}
\includegraphics[width=0.9\linewidth]{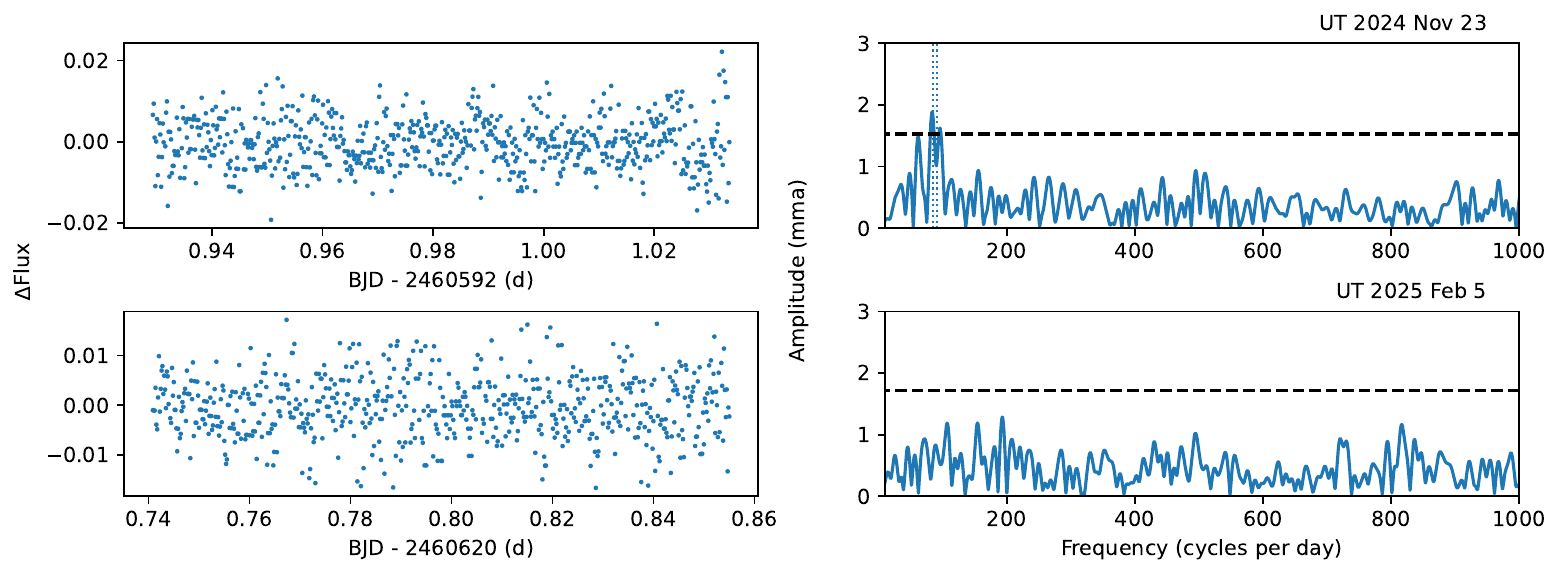}
\caption{Light curves and Fourier transforms for J0912$-$2642 }
\label{fig:j0912}
\end{figure*}

\begin{figure*}
\centering
\vspace{-0.2in}
\includegraphics[width=0.9\linewidth]{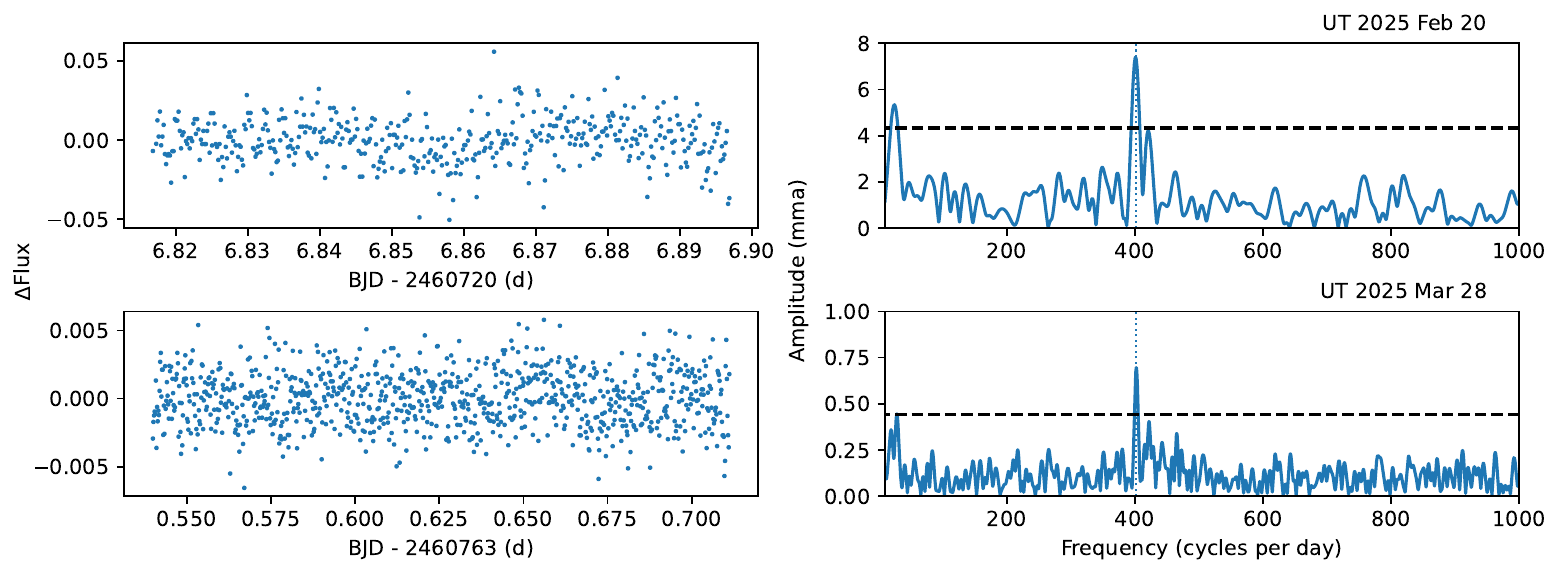}
\caption{Light curves and Fourier transforms for J0959$-$1828. The top and bottom rows show data taken at APO and NTT/ULTRACAM ($g$-band), respectively.}
\label{fig:j0959}
\end{figure*}

\begin{figure*}
\centering
\vspace{-0.2in}
\includegraphics[width=0.9\linewidth]{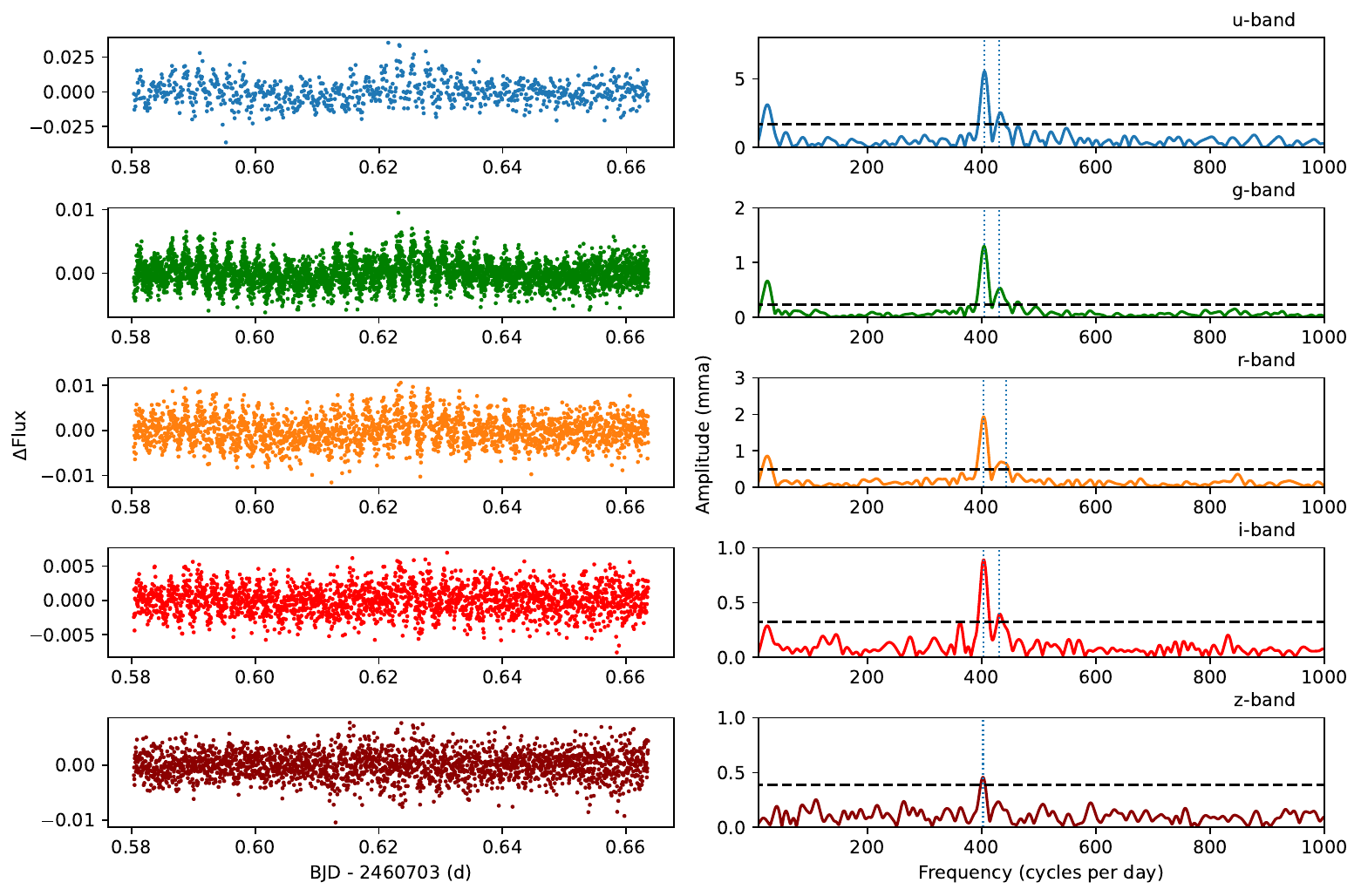}
\caption{Light curves and Fourier transforms for J0959$-$1828 taken at GTC.}
\label{fig:j0959_hiper}
\end{figure*}

\subsection{J0912$-$2642}

J0912$-$2642 has $M=1.262 \pm 0.002~M_\odot$ and $T_{\rm eff}= 12,973 \pm 115$ K. We were able to observe J0912$-$2642 on two nights. On UT 2024 Nov 23, we detected two modes at 85.3 and 91.7 c/d with fairly low amplitudes of 2.3 mma and 1.8 mma, respectively. The second night, UT 2025 Feb 5, revealed no significant modes. Figure \ref{fig:j0912} shows the light curves and Fourier transforms. Given the two modes detected on the first night of observations,
we tentatively classify this as a pulsator. However, additional time-series data are needed to confirm pulsations and study the pulsation spectrum of this
ultra-massive white dwarf.

\subsection{J0959$-$1828}

This white dwarf was previously observed for 5 nights by \citet{kilic23a} at three different observatories. Their observations indicated that
J0959$-$1828 is possibly pulsating, but their data was inconclusive as they detected significant modes above the 4 $\langle{A}\rangle$ limit only
in one of the five nights of observations. 

J0959$-$1828 is an exciting find in our follow up sample, as it has a mass of $M=1.320 \pm0.004~M_\odot$ assuming a CO core. 
Figure \ref{fig:j0959} shows its APO (top) and NTT/ULTRACAM (bottom panels) light curves and Fourier transforms. Our APO observations
reveal a significant mode at $401.2 \pm0.8$ c/d with an amplitude of 7.4 mma. We confirm this pulsation mode using ULTRACAM, and detect a mode at $402.4 \pm0.4$ c/d in
the ULTRACAM $g$-band data, though the amplitude has gone down an order of magnitude. Thanks to ULTRACAM's higher cadence and sensitivity,
this mode with $0.7\pm0.1$ mma is detected significantly above the noise level in the Fourier transform. 
While we also have $u$ and $r$-band data from ULTRACAM, the $u$-band data is noisier, and the pulsations have smaller amplitudes in the red.
We do not detect any significant frequencies in the $u$ or $r$-band, which is not surprising as the $g$-band data only shows low level variability. 

We also observed this white dwarf at GTC. Figure \ref{fig:j0959_hiper} shows the light curves and Fourier transforms. This data further confirms that
J0959$-$1828 is pulsating with significant modes at 404.5 and 430.2 c/d. The former has 1.3 mma in the $g-$band, which goes up to
5.5 mma in the $u-$band. In addition, the second mode at 430.2 c/d is clearly detected in multiple filters, with the highest amplitude
measured in the $u-$band with $2.3\pm0.3$ mma.

In both the data from APO and GTC, there is a low frequency peak, that we tentatively identify as likely real. At APO the peak has a period of 3733.5 s, and at the GTC the peak is at 3421.0 s. These periods are too long to be due to pulsations in DA white dwarfs and no ZZ Ceti exhibits such long pulsation periods. This peak may indicate variability due to rotation, but further observations are required in order to confirm and constrain the source of this peak.

With confirmation of multi-periodic pulsations in this system, J0959$-$1828 becomes the most massive pulsating white dwarf currently known, though
its mass is consistent with the previous record holder J0049$-$2525 within $1\sigma$. J0959$-$1828 is just slightly more massive and about 1000 K cooler.
The variability in J0959$-$1828 is observed at a much lower level compared to J0049$-$2525, making the detection and confirmation of its pulsation
modes more challenging.

\subsection{J1106+1802}
This white dwarf has $M=1.131 \pm 0.011 ~M_\odot$ and $T_{\rm eff} =12,877 \pm 269 K$. It was previously reported as a ZZ Ceti by \citet{guidry}. Their observations revealed three modes at 369.7, 537.2, and 1180.2 s with amplitudes of 10.2, 0.5, and 0.6\%, respectively.

\begin{figure*}
\centering
\vspace{-0.2in}
\includegraphics[width=0.9\linewidth]{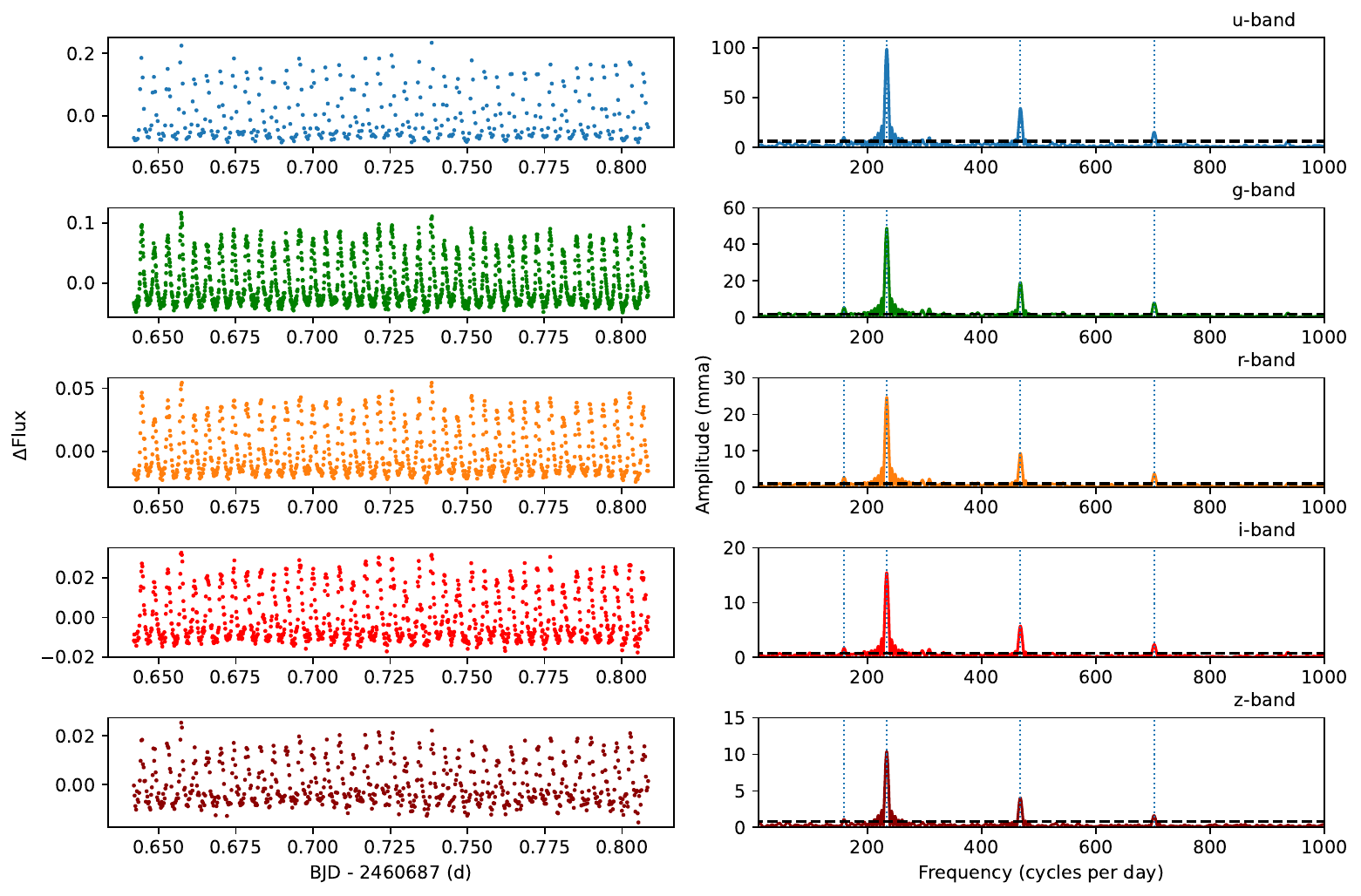}
\caption{Light curves and Fourier transforms for J1106$+$1802 taken on 2025 Jan 12.}
\label{fig:j1106}
\end{figure*}

\begin{figure*}
\centering
\vspace{-0.2in}
\includegraphics[width=0.9\linewidth]{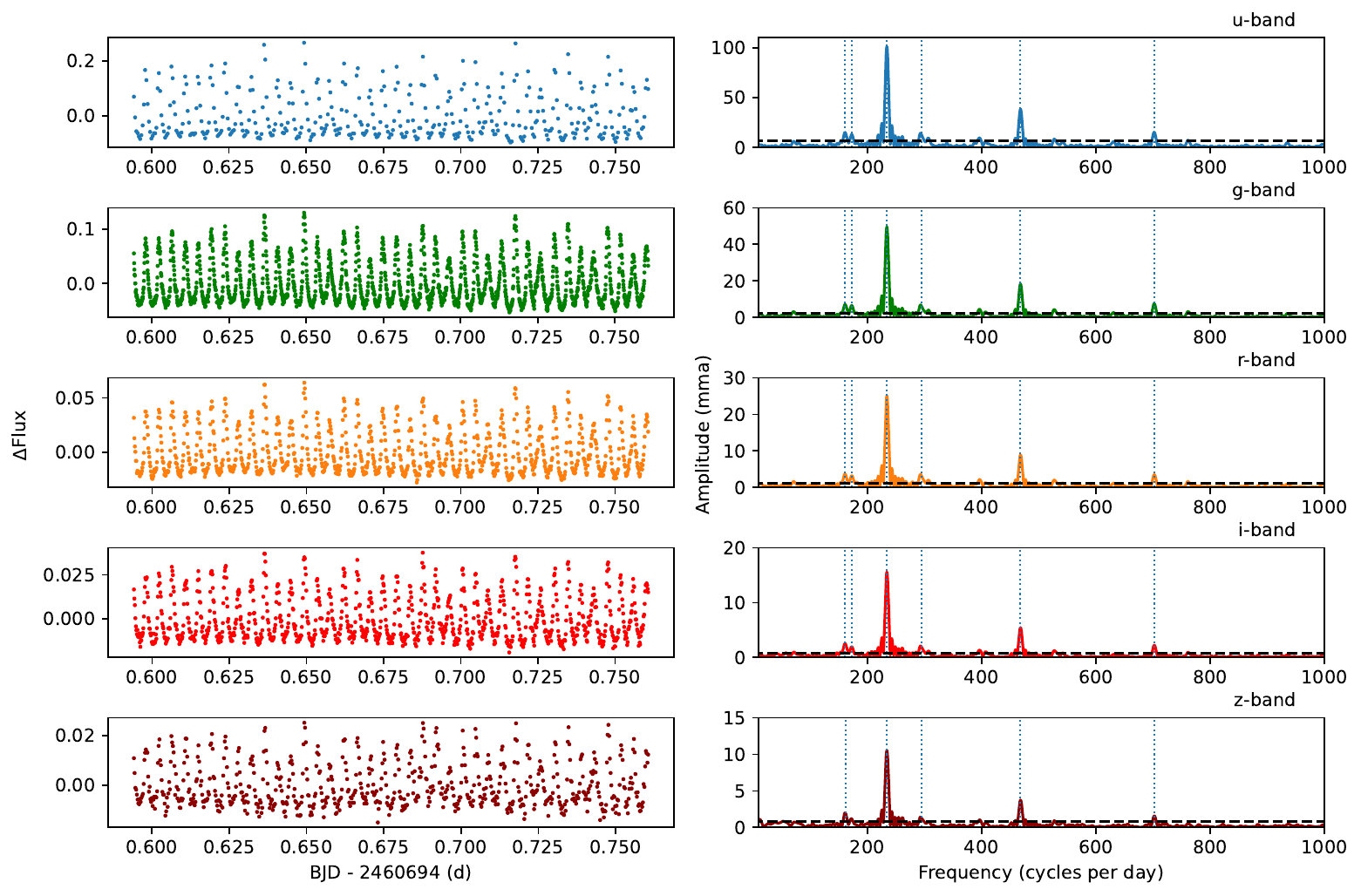}
\caption{Light curves and Fourier transforms for J1106$+$1802 taken on 2025 Jan 19. }
\label{fig:j1106a}
\end{figure*}

We observed J1106+1802 on two nights using HiPERCAM. We include all 5 bands of data in Figure \ref{fig:j1106}, Figure \ref{fig:j1106a}, and Table \ref{tab:freqs_hiper}. On 2025 Jan 12, we find 4 significant modes at 159.2, 234.0, 468.0, and 702.1 c/d that are visible in all filters. However,
the last two are clearly harmonics (twice and thrice the frequency) of the dominant mode at 234.0 c/d, which has an amplitude of $\approx100$ mma in the
$u-$band and $\approx50$ mma in the $g-$band. The second night of GTC observations on 2025 Jan 19 reveals 6 pulsation modes
with frequencies 161.3, 173.4, 234.2, 295.0, 468.0, and 702.1 c/d in the $g-$band. The mode at 234.2 c/d has the highest amplitude again at 49.6 mma, and
latter two modes are clearly its harmonics. The peak at $\approx$ 234 c/d seems to be dominant in all of the data presented here and in \citet{guidry}.

\subsection{J1626+2533}

J1626+2533 was previously classified as an NOV by \citet{vincent}, where they ruled out variability at the $\geq5.8$\% level. 
It has $M=1.140 \pm 0.007~M_\odot$ and $T_{\rm eff} = 13,202 \pm 189$ K.

\begin{figure*}
\centering
\vspace{-0.2in}
\includegraphics[width=0.9\linewidth]{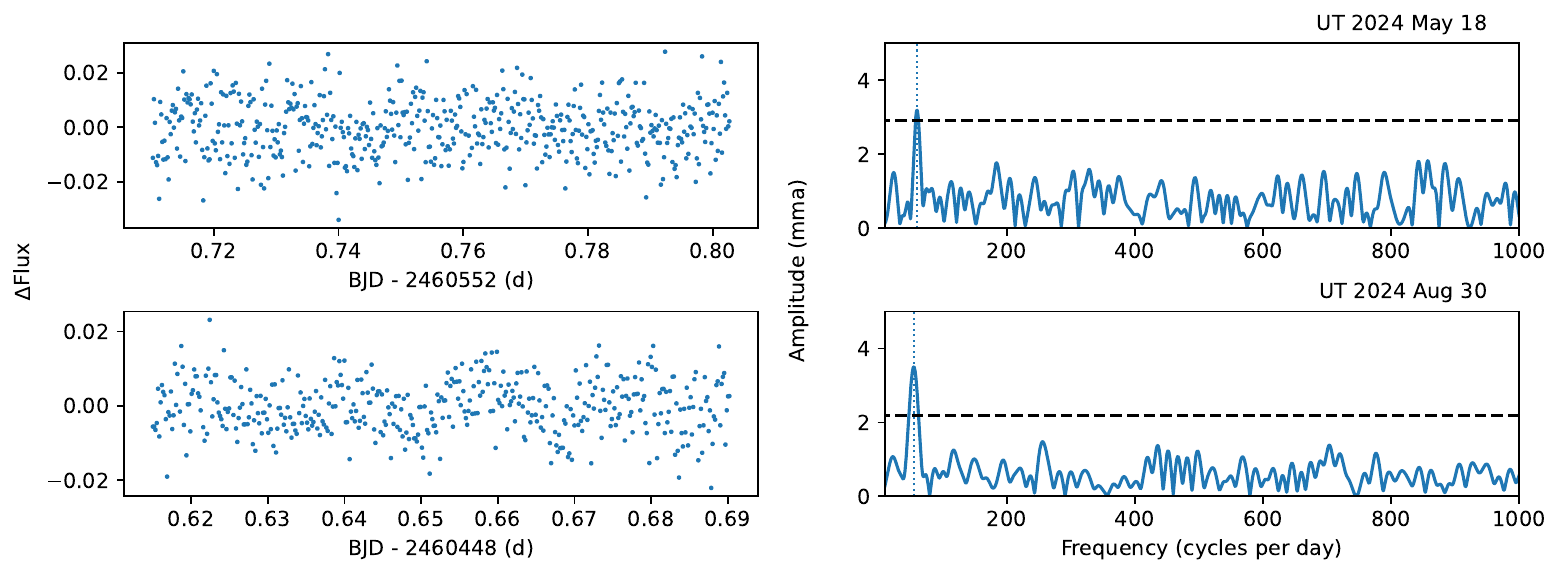}
\caption{Light curves and Fourier transforms for J1626$+$2533 }
\label{fig:j1626}
\end{figure*}

We observed J1626+2533 on UT 2024 May 18 and detected a significant mode at 59.8 c/d with an amplitude of 3.2 mma. On UT 2024 Aug 30, we detected
another significant mode at 55.2 c/d with amplitude 3.6 mma. The light curves and Fourier transforms are shown in Figure \ref{fig:j1626}. 
Hence, we confirm multi-periodic oscillations in J1626+2533, and reclassify it as a pulsating white dwarf. The rest of the newly discovered
ZZ Cetis in our sample are presented in Appendix~\ref{appendix:a}.

\section{Non-variable White Dwarfs}

\citet{jewett2024} reported 14 white dwarfs classified as NOVs in their sample based on the literature data. We re-observed five white dwarfs that were
previously classified as non-variable by \citet{vincent} including: J0408+2323, J0538+3212, J0657+7341, J1243+4805, and J1928+1526. These all lie within
the boundaries of the instability strip, yet we also do not detect any significant variations in these systems. In addition, we find 10 additional NOVs
among the newly observed systems. A summary of these newly identified and previously known NOVs can be found in Table \ref{tab:puls}.

We present the light curves and the corresponding Fourier transforms for the NOVs in Figure \ref{fig:nov} in Appendix \ref{appendix:b}. 
We summarize our APO observations, including the length of observation and 4$\langle {\rm A}\rangle$ significance levels in Table \ref{tab:nov} in Appendix \ref{appendix:b}. 
The detection limits range from 1.5 to 7.7 mma, with a median at 3.1 mma. 

None of these objects show any significant modes above the 4$\langle {\rm A}\rangle$ limits. We were able to observe four of these targets on two separate
nights at APO, and both nights of data confirm the NOV status for those objects. Furthermore, we were able to observe two of these targets, J0408+2323 and
J0657+7341 at GTC using HIPERCAM (see Table \ref{tab:nov_hiper} in Appendix \ref{appendix:b}), with detection limits down to 0.5 mma in the $g-$band, and still did not detect any
significant variability. Because they fall near or within the ZZ Ceti instability strip, it is somewhat
surprising to find so many NOVs in the instability strip. 

Note that \citet{vincent} found a significant discrepancy between the photometric and spectroscopic temperatures for one of these objects, J1243+4805,
where the spectroscopic temperature places it outside of the instability strip. This shows a potential caveat for NOVs close to the edges of the
instability strip. \citet{vincent} suggested that the photometric temperatures might sometimes be underestimated. We discuss the purity of the instability
strip in Section \ref{sec:purity}, and discuss other potential reasons for why we are finding a relatively large number of NOVs in the instability strip
for massive white dwarfs.

\section{Discussion}

\subsection{Trends: Period versus Temperature}

As a DA white dwarf cools, it is predicted to start pulsating when it is cool enough to have a hydrogen partial ionization zone deep enough to
excite global pulsations. With further cooling, the base of the partial ionization zone moves deeper, and the thermal timescale increases, which
leads to the excitation of modes with longer pulsation periods \citep{mukadam,montgomery05}. In addition, the total pulsation power is expected to increase with increasing
mass of the partial ionization zone.

\citet{clemens93} found a trend between the weighted mean period (WMP), where each period is weighted by the corresponding amplitude, and the effective
temperature of a dozen pulsating, average mass DAV white dwarfs: hotter stars have the shortest periods, and the pulsation periods lengthen by
an order of magnitude from the blue edge to the red edge of the instability strip. \citet{clemens93} also detected three orders of magnitude
increase in the total pulsation power over the same temperature and frequency range. The same trends were confirmed with larger samples of average
mass DAVs by \citet{kanaan02} and \citet{mukadam}.

\begin{figure}
\includegraphics[width=3.5in, clip, trim={0in 0in 0in 0in}]{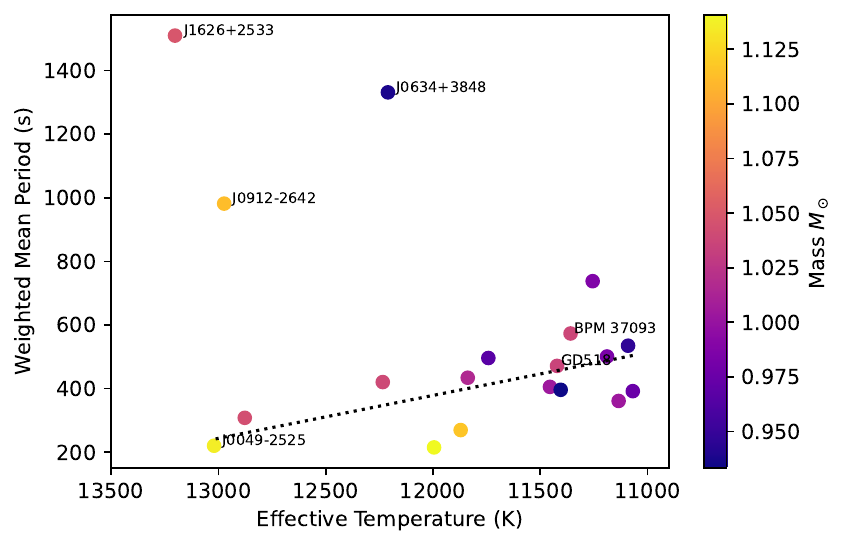}
\caption{Weighted mean period versus effective temperature for the ZZ Ceti white dwarfs in our sample. The color of the point indicates the mass of
each star. We include the well-known ultra-massive ZZ Cetis BPM 37093, GD518, and J0049$-$2525 for reference. The dashed black line shows
the best fitting line to the data, excluding the three outliers: J0634+3848, J0912$-$2642, and J1626+2533, which are labeled.} 
\label{fig:wmp}
\end{figure}

Figure \ref{fig:wmp} shows the WMP versus effective temperature for our sample of massive DAVs, where the color of the points represent the mass of each
object. For targets that were observed on multiple nights, we calculated a single WMP using all of the significant periods from all of the nights,
excluding the harmonics and combination frequencies. We include the previously studied ultra-massive pulsators BPM 37093, GD518, and J0049$-$2525 for reference. BPM 37093 is not in the \citet{jewett2024} sample as it lies outside of the Pan-STARRS footprint. Table \ref{wpmsqrtp} lists the WMPs for each star.

As in previous studies that looked into the relation between the WMP and effective temperature for average mass white dwarfs mentioned above, Figure
\ref{fig:wmp} reveals a clear trend between the WMP and $T_{\rm eff}$ for massive DAVs: WMP increases with decreasing effective temperature. This is mainly because the convection zone deepens into the star. Note also that the cooling of white dwarfs induces an  intrinsic lengthening of the pulsation periods, because the core is increasingly degenerate and the Brunt-V\"ais\"al\"a frequency is smaller \citep[equation 34 in][]{1990ApJS...72..335T}. Finally, cooler white dwarfs of a given mass have a more crystallized core where $g$ modes cannot propagate \citep{1999ApJ...526..976M}, which leads to a smaller propagation region and further lengthening of the period spacing and the periods themselves \citep[see equations 43 and 44 in][]{1990ApJS...72..335T} as clearly depicted in Figure 6 of \cite{2019A&A...621A.100D}. The dashed line of Figure
\ref{fig:wmp} shows a linear fit to the data for comparison. Note that we are not claiming a linear relation between WMP and $T_{\rm eff}$ \citep[see for example][]{clemens93}, this line is simply used to guide the eye.
The previously known ultra-massive DAVs, BPM 37093, GD518, and J0049$-$2525, all follow this trend as well.

\begin{deluxetable}{crr}
\tablecolumns{3} \tablewidth{0pt} 
\caption{Weighted mean period (WMP) and square root of total power.}
\label{wpmsqrtp}
\tablehead{\colhead{Object Name} & \colhead{WMP}  & \colhead{$\sqrt{\rm Total~Power}$}\\
 & (s) & (mma)}
\startdata 
J0039$-$0357 & 269.8 & 66.9\\
J0154$+$4700 & 434.3 & 24.5\\
J0158$-$2503 & 420.6 & 60.3\\
J0204$+$8713 & 361.3 & 22.2\\
J0634$+$3848 & 1331.1 & 9.0\\
J0712$-$1815 & 496.3 & 38.7\\
J0912$-$2642 & 981.5 & 2.9\\
J0959$-$1828 & 214.4 & 7.4\\
J1052$+$1610 & 737.7 & 6.9\\
J1106+1802 & 308.3 & 24.3\\
J1451$-$2502 & 501.0 & 25.7\\
J1626$+$2533 & 1509.5 & 4.8\\
J1722$+$3958 & 392.1 & 24.2\\
J1929$-$2926 & 405.6 & 12.3\\
J2026$-$2254 & 396.3 & 37.6\\
J2208$+$2059 & 534.9 & 10.1 
\enddata
\end{deluxetable}

A striking revelation in Figure \ref{fig:wmp} is that there are three outliers: J0634$+$3848, J0912$-$2642, and J1626$+$2533, which are labeled on
the plot. J1626+2533 is the hottest variable in our sample, yet it has the longest WMP, which is the exact opposite of the visible trend in this
diagram. J1626+2533 was observed on two different nights, with each night of data revealing only 1 significant mode, but those modes have
periods differing by $>3\sigma$. Hence, they are unlikely to be from rotation. The multi-periodic variability demonstrates that J1626+2533 is a
pulsator. It is possible that the relatively long period oscillations observed so far in this system with periods of 1445.5 and 1566.4 s only represent
part of the power spectrum of this object, and additional time-series photometry is required to understand its nature. Further support for this comes
from another outlier, J0634+3848, which shows both short period and long period variability. The photometric data for J0634+3848, see Figure \ref{fig:j0634}
in the Appendix, reveals variability at both $\sim100$ s and $\sim$2,100 s. Since the latter have larger amplitudes, they dominate the WMP measurement.
Hence, it is likely that our WMP measurement may not represent the actual period distribution in this star. 

The last outlier visible in Figure \ref{fig:wmp}, J0912$-$2642, is a bit of a mystery. We were only able to detect multi-periodic variations (two modes)
with amplitudes $\sim2$ mma in this star once, but we only have two nights of observations. It is possible that the relatively low amplitudes seen
in this ultra-massive white dwarf are similar to J0959$-$1828, and that this white dwarf also requires additional observations to confirm
the pulsation modes and constrain its WMP accurately. 

\subsection{Trends: Period versus Mass}

\begin{figure}
\includegraphics[width=3.5in, clip, trim={0in 0in 0in 0in}]{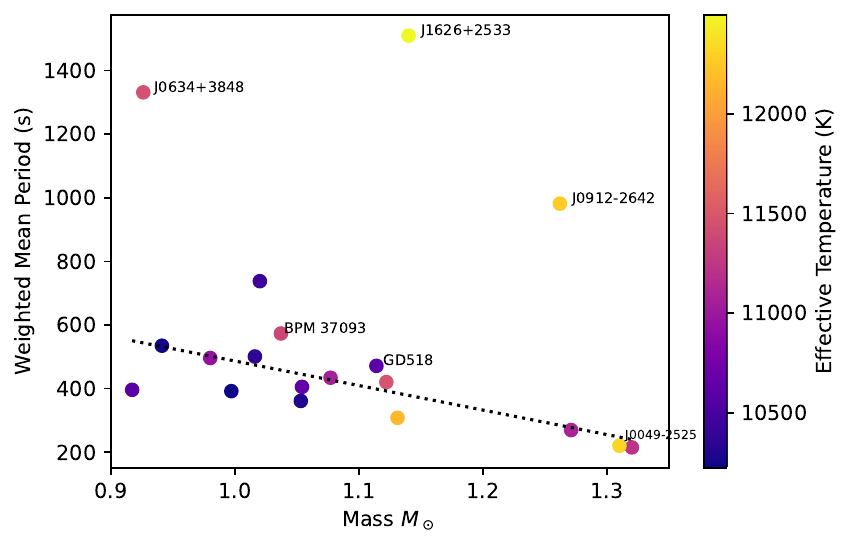}
\caption{Weighted mean period versus mass for the ZZ Ceti white dwarfs in our sample. The color of the points are representative of the effective
temperature of the star. The dashed black line shows the best fitting line to the data, excluding the three outliers.} 
\label{fig:wmpmass}
\end{figure}

Theoretical models predict a dramatic change in pulsation periods with mass. The larger the mass, the higher the gravity, and therefore the higher the Brunt-V\"{a}is\"{a}l\"{a} frequency, see equation 34 in \citet{1990ApJS...72..335T}. This shifts the entire spectrum of $g$-mode periods towards shorter periods. The predicted trend can be seen, for instance, in Figure 9 in \citet{bradley} for ZZ Ceti stars, and in Figure 6 in \cite{2019A&A...621A.100D}  for ultra-massive ZZ Ceti stars. Hence, periods and period spacings are expected to decrease with increasing mass.

A comparison between the pulsation periods for extremely low-mass (ELM) and average mass DAVs shows that the variability timescale is
significantly longer for ELM white dwarfs \citep{hermes13}. For example, the ELM white dwarf companion to the milli-second pulsar PSR J1738+0333
shows pulsation modes with periods ranging from about 1,800 s to 4,980 s, and potentially to even longer periods \citep{kilic18}. Besides this comparison between
ELM white dwarfs and average mass DAVs, we have not been able to find an observational study of the correlation between mass and pulsation period for large
samples of pulsating white dwarfs in the literature.

For ultra-massive white dwarfs, crystallization also needs to be considered because for a given effective temperature, the crystallized mass fraction
in the core is larger for higher masses. Since $g$-modes cannot propagate through the crystallized regions, this reduces the region where the $g$-modes
can propagate. Therefore, for massive white dwarfs with crystallized cores and a reduced $g$-mode propagation zone, period spacing is larger and periods
tend to become longer, see equations 43 and 44 in \citet{1990ApJS...72..335T} and 
Figure 6 of \cite{2019A&A...621A.100D}. Crystallization has the opposite effect of a larger Brunt-V\"{a}is\"{a}l\"{a} frequency, but
the latter is dominant and the overall expectation is that high mass white dwarfs should display shorter pulsation periods.

Figure \ref{fig:wmpmass} shows the WMP versus mass for our massive pulsating DAV sample. The dashed line shows a linear fit to the data to guide the eye.
As expected from the pulsation models, the WMP indeed decreases with increasing mass, from about 600 s for $0.9~M_\odot$ white dwarfs to
$\sim$200 s for the most massive pulsators at $1.3~M_\odot$. This is a nice confirmation of the predicted trend with mass for massive ZZ Ceti white
dwarfs. Not surprisingly, the three outliers in the WMP versus $T_{\rm eff}$ plane, J0634$+$3848, J0912$-$2642, and J1626$+$2533, are also
outliers in this diagram. But the three previously studied ultra-massive pulsators, BPM 37093, GD 518, and J0049$-$2525, 
fall in nicely with our sample. In fact, the WMP of J0049$-$2525 is very close to the newly confirmed ultra-massive pulsator J0959$-$1828.   

\subsection{Trends: Period versus Power}

Studying a sample of $\sim$100 average mass DAVs, \citet{mukadam} found a correlation between the square root of total power\footnote{Total power is defined as the sum of the amplitudes squared.} versus effective temperature or WMP, since WMP is correlated with $T_{\rm eff}$ \citep[also see][]{clemens93}. They found that once the pulsations set in at the blue edge of the
instability strip, the total power increases with decreasing temperature, reaches a maximum in the middle of the instability strip (around WMP
$\sim800$ s), and then declines near the red-edge, just before the pulsations shut down.

\begin{figure}
\includegraphics[width=3.5in]{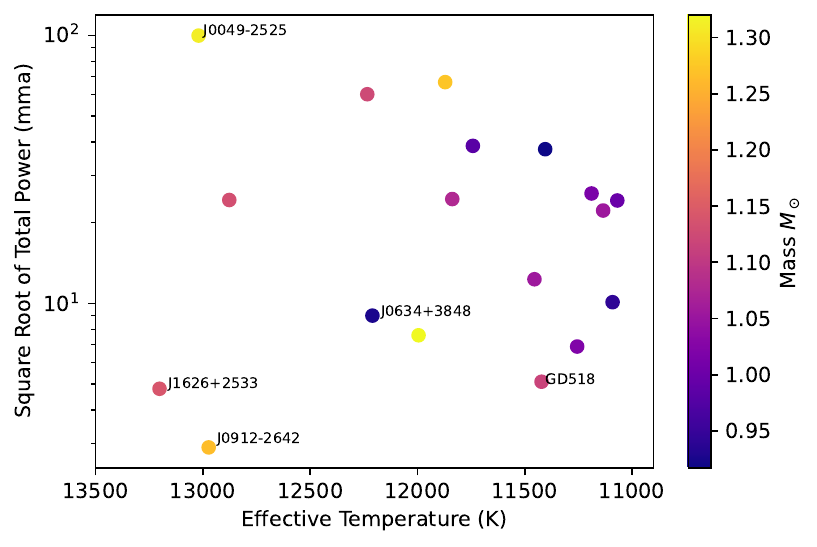}
\caption{Square root of the total power versus effective temperature for the observed pulsators in our sample plus GD 518 and J0049$-$2525. The y axis is set to a log scale, and the color of the points are indicative of the white dwarf's mass. The three outliers from Figure \ref{fig:wmp} are also labeled.}
\label{fig:srtp}
\end{figure}

Figure \ref{fig:srtp} shows the square root of the total power versus effective temperature for our sample of massive DAV white dwarfs, with each
point's color corresponding to its mass. Table \ref{wpmsqrtp} lists the square root of the total power for each star for reference.  
Our sample is much smaller than the one presented in \citet{mukadam}, but we confirm the overall trend seen in average mass DAVs.
The pulsators in the middle of the instability strip on average show pulsations with the highest power, and the total power clearly decreases towards
the red edge. 

The blue edge of the instability strip is more complicated. Even though J0049$-$2525 is one of the hottest pulsators in our sample, it shows
the highest total power, but it's also one of the most massive pulsators known with a significantly crystallized core \citep{kilic23b,2025ApJ...988...32C}.
On the other hand, the two other objects near the blue edge, J0912$-$2642 and J1626+2533, have significantly lower total power, which would be more
in line with lower amplitudes expected near the blue-edge. However, these two stars are also outliers in Figure \ref{fig:wmp}. Hence, we cannot
constrain the overall trend in total power for this massive white dwarf sample without additional follow-up observations to constrain the nature
of these outliers.

\subsection{The Purity of the ZZ Ceti Strip} \label{sec:purity}

\begin{figure*}
\centering 
\includegraphics[width=0.75\linewidth]{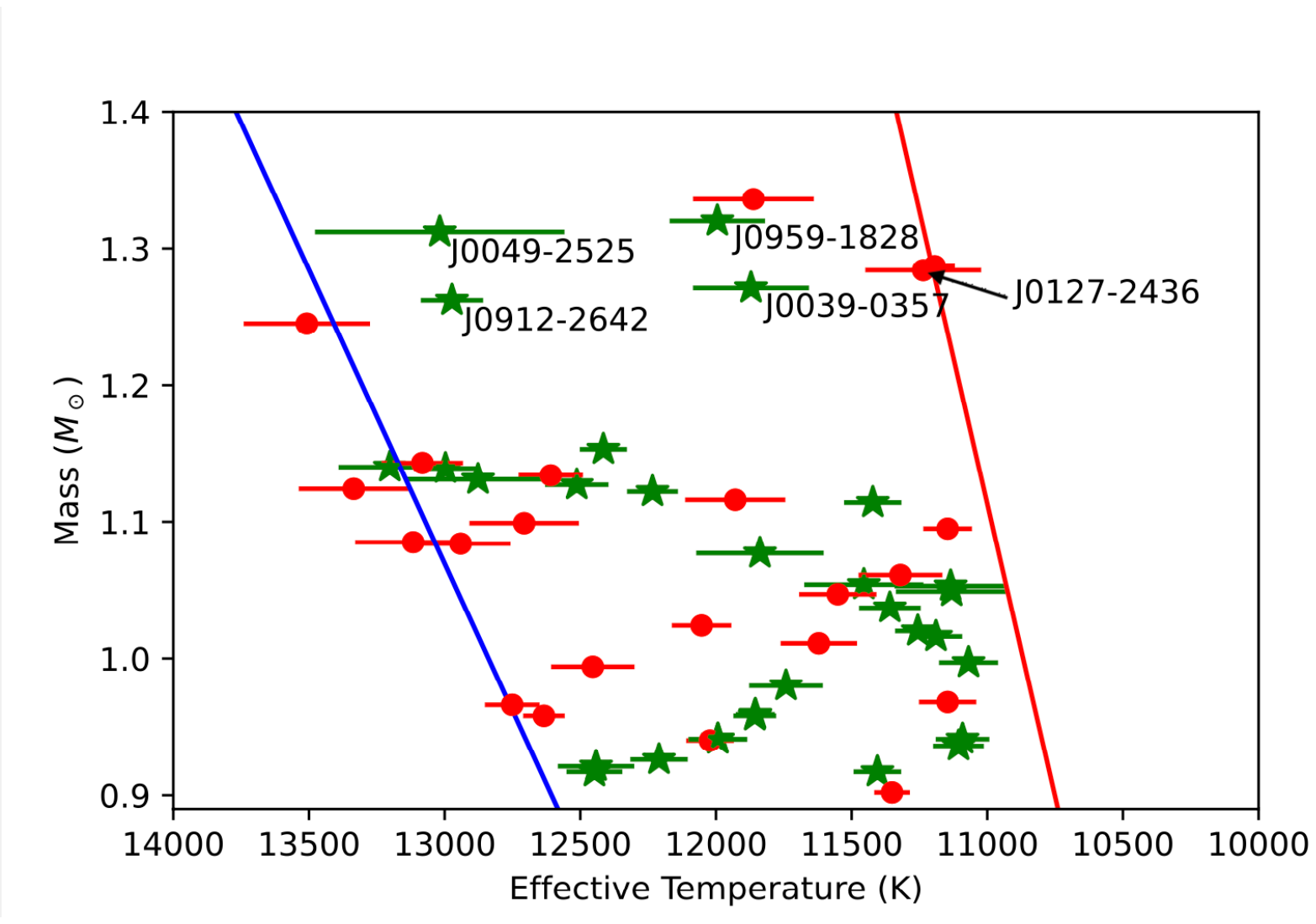}
\caption{Masses and effective temperatures for our follow up sample of massive DA white dwarf sample including white dwarfs from this work and those previously known in the literature and the four most massive pulsators labeled and J0127$-$2436 which lies on the edge of the instability strip. The objects marked with by a red point are NOVs. The green stars mark the pulsators. The blue and red lines show the empirical boundaries of the photometric ZZ Ceti instability strip from \citet{vincent}.}
    \label{fig:zzceti}
\end{figure*}

\begin{deluxetable}{cccccc}
\tablecolumns{6} \tablewidth{0pt}
\caption{Massive and ultra-massive pulsators and NOVs in MWDD 100pc sample.} \label{tab:puls}
\tablehead{\colhead{Pulsators} & \colhead{Mass ($M_\odot$)} & \colhead{Reference} & \colhead{NOVs} & \colhead{Mass ($M_\odot$)} & \colhead{Reference}}
\startdata
J0039$-$0357 & 1.271 $\pm{0.009}$&This Work & J0050$-$2826 &1.061 $\pm{0.011}$&This Work\\
J0049$-$2525 &1.312 $\pm{0.010}$&\citet{kilic23b} & J0127$-$2436 & 1.284 $\pm{0.009}$&This Work \\
J0135$+$5722 &1.153 $\pm{0.004}$&\citet{degeronimo} &  J0135$+$5722 &1.153 $\pm{0.004}$&\citet{vincent}\\
J0154$+$4700 &1.077 $\pm{0.014}$&This Work& J0234$-$0511 &0.958 $\pm{0.004}$&\citet{gianninas}\\
J0158$-$2503 & 1.122 $\pm{0.007}$&This Work& J0347$-$1802 &1.143 $\pm{0.007}$&\citet{guidry}\\
J0204$+$8713 & 1.053 $\pm{0.015}$&\citet{vincent}  \& \citet{jewett2024} \& This Work & J0408$+$2323 & 1.024 $\pm{0.008}$&\citet{vincent} \& This Work\\
J0448$-$1053 &0.941 $\pm{0.007}$&\citet{romero} & J0538$+$3212 &0.994 $\pm{0.010}$&\citet{vincent} \& This Work \\
J0551$+$4135 &1.139 $\pm{0.005}$&\citet{vincent} and M. Uzundag et al. (submitted)  & J0634$+$3848 &0.926 $\pm{0.006}$&\citet{vincent}\\
J0634$+$3848&0.926 $\pm{0.006}$&This Work& J0657$+$7341 &1.134 $\pm{0.007}$&\citet{vincent} \& This Work\\
J0712$-$1815&0.980 $\pm{0.008}$&This Work& J0725$+$0411 &0.940 $\pm{0.008}$&This Work\\
J0856$+$6206 &0.958 $\pm{0.007}$&\citet{vincent} & J0949$-$0730 &1.084 $\pm{0.008}$&This Work\\
J0912$-$2642&1.262 $\pm{0.002}$&This Work& J0950$-$2841 &1.124 $\pm{0.006}$&This Work\\
J0959$-$1828&1.320 $\pm{0.004}$&This Work& J1140$+$2322 &1.336 $\pm{0.006}$&\citet{kilic23a}\\
J1052$+$1610&1.020 $\pm{0.009}$&This Work& J1243$+$4805 &0.966 $\pm{0.006}$&\citet{vincent} \& This Work\\
J1106$+$1802 &1.131 $\pm{0.011}$&\citet{guidry} \& This Work & J1342$-$1413 &0.902 $\pm{0.006}$&This Work\\
BPM 37093 &1.037 $\pm{0.008}$&\citet{kanaan} & J1552$+$0039 &1.245 $\pm{0.011}$&This Work\\
J1451$-$2502 &1.016 $\pm{0.011}$&This Work& J1626$+$2533 &1.140 $\pm{0.007}$&\citet{vincent}\\
J1626$+$2533 &1.140 $\pm{0.007}$&This Work& J1655$+$2533 &1.287 $\pm{0.002}$&\citet{curd17}\\
GD 518 &1.114 $\pm{0.006}$&\citet{hermes} & J1656$+$5719 &1.047 $\pm{0.010}$&This Work\\
J1722$+$3958 &0.997 $\pm{0.010}$&This Work& J1813$+$4427 &1.095 $\pm{0.007}$&\citet{vincent}\\
J1812$+$4321 &0.921 $\pm{0.007}$&\citet{romero} & J1819$+$1225 &1.116 $\pm{0.012}$&This Work\\
J1929$-$2926 &1.054 $\pm{0.016}$&This Work& J1910$+$7334 &1.085 $\pm{0.008}$&\citet{vincent}\\
J2026$-$2254 &0.917 $\pm{0.008}$&This Work& J1928$+$1526 &1.011 $\pm{0.012}$&\citet{vincent} \& This Work\\
J2208$+$2059 &0.941 $\pm{0.011}$&This Work& J2107$+$7831 &1.099 $\pm{0.009}$&This Work\\
\enddata
\tablecomments{Masses from \citet{jewett2024}. BPM 37093 mass from \citet{obrien} and is outside of the MWDD 100 pc sample, but is included for reference.}
\end{deluxetable} 

Figure \ref{fig:zzceti} shows the pulsators (green) and NOVs (red points) in the MWDD 100 pc sample along with the empirical boundaries of the photometric instability
strip from \citet{vincent}. The four most massive pulsators are labeled. This plot of the massive MWDD 100 pc sample shows a much more
complete picture of the massive and ultra-massive pulsators compared to the one presented in \citet{jewett2024}, as now every white dwarf that is near or 
within the boundaries of the instability strip has time-series photometry available. For reference, the most massive white dwarf on this plot is
J1140+2322, which was classified as a NOV by \citet{kilic23a}. 

There is a large number of NOVs within the instability strip. It is possible
that the slope of the red edge could be slightly different, which could help remove some of the offending NOVs from the sample. However, we do not try
to adjust this boundary based on our observations, since observations of additional white dwarfs lying outside of the strip would be necessary to determine
the exact location of the red edge for these massive white dwarfs. 

The purity of the ZZ Ceti instability strip has been studied extensively in the literature. If it is pure, if every star in the instability strip
shows pulsations, this would indicate that this is an evolutionary phase that every DA white dwarf goes through. Hence, we can use the internal
structures derived through asteroseismology, such as constraints on the surface H layer thickness, to the entire population of DA white dwarfs.
In addition, if the strip is pure, it also enables identification of additional pulsators solely through measurements of their atmospheric parameters
\citep{gianninas05}. 

In their search for DAV white dwarfs within the SDSS spectroscopy sample, \citet{mukadam04} found several non-variables within the instability strip,
and questioned its purity. \citet{cas07} followed up two of these non-variables, and found them to show low-amplitude variability that escaped
detection in earlier observations. They suggested that the ZZ Ceti strip is likely pure, but also highlighted other NOVs that needed
to be observed. \citet{bergeron04} found that the spectroscopic method for estimating the atmospheric parameters had a
100\% success rate for predicting which stars should pulsate. Their analysis found no NOVs within the ZZ Ceti strip, supporting the idea that the instability
strip is pure and that all DA white dwarfs evolve through this phase. 

We find answering the question of whether all DA white dwarfs evolve through the ZZ Ceti phase irrelevant. Obviously, all white dwarfs will pass through the
instability strip as they cool. The actual question is: what are the potential mechanisms that can stop or hide pulsations in these white dwarfs as
they go through the strip? There are several candidates:

1- Magnetic fields have been suggested to suppress pulsations in white dwarfs by affecting atmospheric convection, thus disrupting the driving
mechanism of pulsations. \citet{tremblay} demonstrated that magnetic fields stronger than about 50 kG are sufficient to suppress convection in
white dwarf photospheres, which was later confirmed through observations of WD2105$-$820, where only the radiative models can explain the spectral
energy distribution of this weakly magnetic white dwarf \citep{gentile}. Hence, magnetism is likely to stop pulsations in DA white dwarfs.
Low resolution spectroscopy is not sensitive to fields below 100 kG for massive DA white dwarfs \citep{kilic2015}. Hence, such weakly magnetic stars
would appear as normal DAs, yet would not show pulsations. 

\citet{bagnulo22} found that average mass white dwarfs are rarely magnetic at birth, but weak fields appear slowly, especially after the stars go
through core crystallization. For example, out of the 100 objects with $0.5<M<0.8~M_\odot$ and $T_{\rm eff}\geq10,000$ K in their sample, only
4 are magnetic: WD 1105$-$048 (0.01 MG), WD 1105$-$340 (0.15 MG), WD 1704+481.1 (0.01 MG), and WD 2047+372 (0.06 MG). Even though magnetism
can stop pulsations in white dwarfs, it is rare to find magnetism in average mass white dwarfs while they are going through the ZZ Ceti strip, and therefore
we would expect the instability strip to be nearly ($\sim96$\%) pure for those most common DAVs.

On the other hand, massive white dwarfs,
a large fraction of which are likely merger remnants \citep{temmink,jewett2024}, tend to be strongly magnetic at birth. For comparison,
out of the 15 objects with $M\geq0.9~M_\odot$ and $T_{\rm eff}\geq10,000$ K in the \citet{bagnulo22} sample, 9 are magnetic.
Only one of the massive magnetic white dwarfs in that sample, WD 2051$-$208 ($B=0.3$ MG) has a field strength lower than 1 MG.
\citet{gianninas} classified this white dwarf as a DA based on a low-resolution spectrum that covers only up to H$\beta$. Hence,
we can assume a 1 in 7 contamination rate of the ``apparently'' non-magnetic massive DA sample by weakly magnetic stars like WD 2051$-$208.
Using a binomial distribution, this corresponds to a fraction of $14.3_{-7.1}^{+15.1}$\%. Hence, the ZZ Ceti strip for massive white dwarfs
could be contaminated by weakly magnetic DAs where the field strength is not strong enough to cause visible Zeeman split lines in low-resolution
spectra obtained by \citet{jewett2024}. This could explain some of the NOVs in our sample.

\begin{figure*}
\centering
\includegraphics[width=2in]{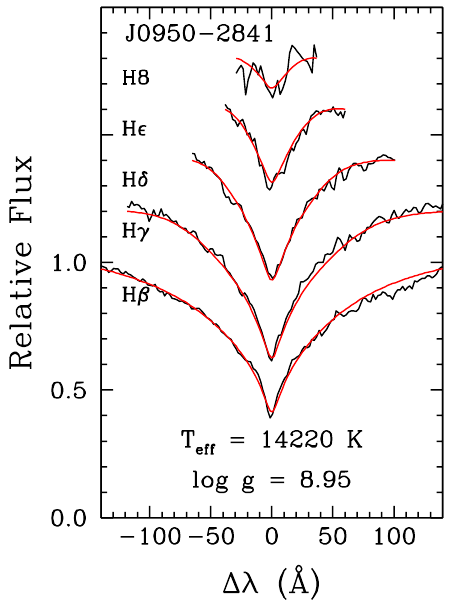}
\includegraphics[width=2in]{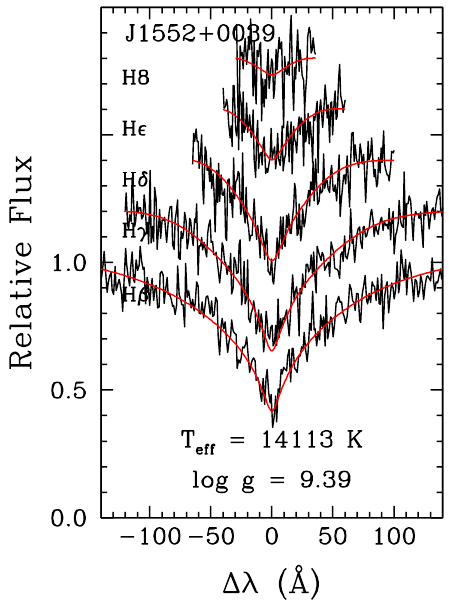}
\includegraphics[width=2in]{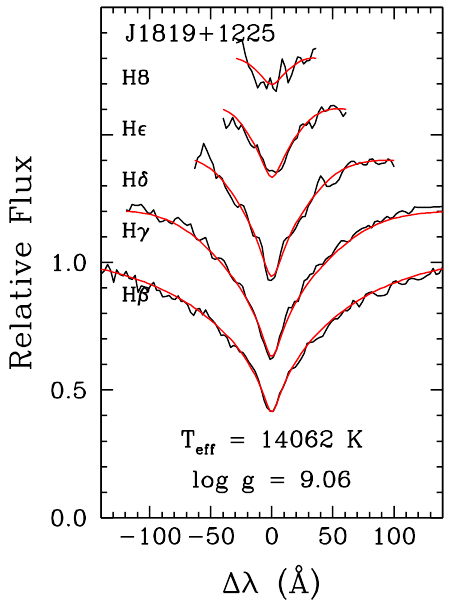}
\caption{Spectroscopic model fits (red) to observed Balmer line profiles (black) of three NOV white dwarfs. The best-fitting effective temperature and $\log{g}$ are shown at the bottom of each panel. The effective temperatures and $\log{g}$ values have been corrected for 3D hydrodynamical effects \citep{tremblay3d}.}
\label{fig:specfits}
\end{figure*}

2- Given the relatively low-amplitude pulsations observed in several massive white dwarfs in our sample, our precision may not be sufficient to detect pulsations
in all. J0959$-$1828 is an excellent example, where the pulsation amplitude changed by an order of magnitude over the three nights of observations, and
we could confirm pulsations in this system only because of the exquisite data from ULTRACAM and HIPERCAM. Without those additional data, we would not be
able to detect the $\sim0.7$ mma oscillations in this target. Our sensitivity limits for the NOVs presented in Table \ref{tab:nov} range from 1.5 to 7.7 mma,
whereas the two NOVs observed with HIPERCAM (see Table \ref{tab:nov_hiper}) have detection limits down to 0.5 mma in the $g-$band. 
Hence, we cannot rule out low-amplitude pulsations in a significant fraction of the NOVs that we identified. 

Based on these arguments, additional time-series observations of the NOVs in our sample would be helpful to search for low amplitude pulsations and
confirm their nature. However, given the relatively large fraction of magnetic white dwarfs among the young and massive white dwarfs in volume-limited
samples \citep{bagnulo22,jewett2024}, perhaps it is not surprising that some of these massive white dwarfs are weakly magnetic. Such objects could
appear as normal DAs in low-resolution spectra, yet may not show any variability while they go through the ZZ Ceti instability strip.

3- The method in which the physical parameters of the white dwarf are calculated can impact the position of the star in the instability strip,
potentially pushing it outside of the bounds. There are two methods that are generally used to estimate the physical parameters of white dwarfs: the
spectroscopic and photometric methods. The parameters used in our study are based on the photometric method \citep{jewett2024}.
For one of the NOVs in our sample, J1243+4805, \citet{vincent} report a spectroscopic temperature of 14\,838 K, which is much hotter than
the photometric temperature of 12\,751 K and would push this NOV beyond the blue edge of the instability strip. 

The spectroscopic method requires accurate flux calibration and high signal-to-noise ratio spectra covering the gravity-sensitive higher order Balmer
lines. The available spectra for the majority of our targets do not have sufficient signal-to-noise ratio for such an analysis, but we identify at least
five NOV white dwarfs  that are outside of the ZZ Ceti strip based on their spectroscopic temperatures.
Figure \ref{fig:specfits} shows our spectroscopic fits to three of these NOVs, where the spectroscopic temperatures are all hotter than
14\,000 K, which would push all three stars outside of the instability strip. Either the photometric method is underestimating or the spectroscopic
method is overestimating their temperatures by 10-20\%. This is consistent with \citet{GB}, who found that the spectroscopic temperatures of DA stars
exceed the photometric values by $\sim10$\% above 14\,000 K likely due to inaccurate treatment of Stark broadening in the model spectra. Other factors affecting the spectroscopic temperature in this range include atmospheric convection, 3D hydrodynamical effects, and flux calibration \citep{GB}.

%This shows evidence that the method used to estimate the temperature of the white dwarf can impact whether an object is located
%within the ZZ Ceti strip. 
On the other hand, there are other NOVs where the spectroscopic and photometric temperatures agree. For example, 
for J0127$-$2436 we find a spectroscopic temperature of $T_{spec} = 11\,557$ K compared to the photometric temperature $T_{phot} = 11\,236$ K. The spectroscopic method does not yield a significantly different temperature in this case, leaving this NOV still within the bounds of the photometric ZZ Ceti strip. It is interesting to note that the red edge of the spectroscopically determined instability strip is about 600 K
hotter at $\approx1.2M_\odot$, potentially placing this star outside of the instability strip. The photometric and
spectroscopic temperatures are in agreement, yet whether the star is within the instability strip or not is unclear. Again, the red edge of the photometric strip possibly needs to be adjusted to exclude this star, but more observations of NOVs outside of the strip are required. While the temperature discrepancy can explain some the of the NOVs in the strip, it cannot account for all of them.  

Typically, massive white dwarfs are assumed to have CO cores and ultra-massive white dwarfs are assumed to have ONe cores. However, the actual composition of the cores in ultra-massive white dwarfs remains uncertain. Ultra-massive white dwarfs are formed either through single stellar evolution with a single progenitor or through binary evolution with a merger event of two white dwarfs \citep{temmink}. Both evolutionary channels can produce ultra-massive white dwarfs with CO and ONe cores, leaving the true nature of ultra-massive white dwarf interiors unresolved (see discussion in \citet{camisassa21}. 

White dwarf mass estimates depend on the assumed core composition. For example, an ONe core would imply a mass lower by $0.05~M_{\odot}$ compared to
the CO-core case for the pulsator J0959$-$1828. The majority of the NOVs in our sample are located well within the instability strip boundaries where
small mass differences would not make a significant difference in the white dwarf's location relative to the boundaries, see Figure \ref{fig:zzceti}.
For the purposes of this study, assuming a CO core composition is sufficient.

4-There is the possibility for some pulsation modes to show no (or very little) variation for some specific geometry, or its possible that destructive interference
between two or more oscillation modes can also lower the pulsation amplitudes below the detection threshold \citep{gianninas}.  Given these caveats, perhaps it is not
surprising that there are NOVs within the instability strip for massive white dwarfs.

\subsection{Potential rotational multiplets}

A well-known characteristic of non-radial stellar pulsations is that the eigenfrequencies of modes with spherical degree $\ell$ may split into $2\ell + 1$ components due to stellar rotation. These components are distinguished by their azimuthal order $m$. Under the assumption of slow and rigid rotation, the frequency splitting is given by
\begin{equation}
\delta \nu_{\ell, k, m} = m \left(1 - C_{\ell,k}\right) \Omega_{\rm R},
\end{equation}
 where $\Omega_{\rm R}$ is the stellar rotation frequency expressed in cyclic units (e.g.,~$\mu$Hz or c/d),
%where $\Omega_{\rm R}$ is the rotational angular frequency of the star, 
$C_{\ell,k}$ is the Ledoux constant, and $m = 0, \pm1, \pm2, \ldots, \pm \ell$ \citep{1989nos..book.....U}. The condition for the validity of this first-order expression requires that $\Omega_{\rm R} \ll \delta\nu_{\ell,k}$. 

In a few of the ultra-massive WDs in our sample, we identify closely spaced frequency multiplets that could in principle be interpreted as signatures of rotation. However, for several stars (e.g., J1106+1802 and J0039$-$0357) the derived rotation frequencies are of the same order as the pulsation frequencies themselves (rotation periods of $\sim$0.2 hr correspond to $\delta\nu_{\ell} \sim 120 $–140 c/d). This violates the slow-rotation condition, and therefore the simple Ledoux splitting expression cannot be applied in a strictly valid way. The frequency spacings we report should thus be regarded as 'candidate multiplets', not secure detections of rotational splitting.  

An additional complication arises from the fact that frequency differences in the range we observe correspond to period differences of $\sim$20–100 s. Such spacings are comparable to the expected asymptotic $g$-mode period spacings for ultra-massive WDs. For ONe-core models in the mass range $1.10$–$1.29\,M_{\odot}$ and effective temperatures within the ZZ Ceti instability strip, theoretical calculations predict asymptotic $\ell=1$ spacings of $\sim$21–39 s \citep[see Fig.~6 of][]{2019A&A...621A.100D}. While some of the observed spacings fall near this range, others are significantly larger, making it difficult to disentangle whether the structures are due to rotational splitting, consecutive radial overtones, or a combination of both effects. Similar cautions have been emphasized by \citet{2015ASPC..493...65K}.  

We summarize the observed frequency patterns in Table~\ref{tab:rotation}. For each star we report the potential doublets, triplets, or higher-order structures and the corresponding mean frequency spacings. While these values could be consistent with rotationally split multiplets, we stress that unambiguous rotation periods cannot be derived from the present data. Additional observations and detailed modeling will be required to distinguish between true multiplets and intrinsic $g$-mode period spacings.
For the remaining ultra-massive white dwarfs in our sample, we do not find compelling evidence for the presence of rotationally split multiplets. This implies that additional data are required to resolve potential multiplet structures. 

\begin{deluxetable}{lcc}[b]
\caption{Summary of potential rotationally split multiplets in ultra-massive white dwarfs presented in this work.\label{tab:rotation}}
\tablehead{
\colhead{Star Name} & \colhead{Multiplet Type} & \colhead{$\langle \delta \nu \rangle$ (c/d) / $P_{\rm rot}$ (hr)}
}
\startdata
J0039$-$0357  & Triplet \& Doublet       & 69.77 / 0.17 \\
J0158$-$2503  & Quadruplet ($\ell=2$)    & 10.40 / 0.25 \\
J0204$+$8713  & Doublet    ($\ell=1$)    & 8.38 / 1.43 \\
J1106$+$1802  & Triplet                  & 60.80 / 0.20 \\
J2208$+$2059  & Doublet    ($\ell=1$)    & 17.85 / 0.67 \\
\enddata
\end{deluxetable}

\section{Conclusions}

Here we present the first time-series photometry of 22 massive white dwarfs from the MWDD 100 pc sample that fall within the ZZ Ceti instability strip. Our full sample contains 31 white dwarfs in which we classify 16 as pulsators and 15 as non-variable. 

Overall, we find four pulsating white dwarfs with $M>1.2~M_\odot$, three of which are new discoveries. Among the new discoveries,
two of the most interesting white dwarfs are J0039$-$0357 and J0959$-$1828. J0039$-$0357 was observed on five nights with different pulsation modes emerging and vanishing, with only one mode consistently found over the five nights. J0959$-$1828, is now the most massive pulsating white dwarf known with mass $1.320 \pm 0.0004~M_\odot$. 

There is a clear correlation between the weighted mean period and temperature; hotter pulsators tend to have shorter periods \citep{clemens93}. Confirmation of this trends comes from other studies in the literature, as well as our sample \citep{kanaan02, mukadam}. We find 3 outliers in our sample: J0634$+$3848, J0912$-$2642, and J1626$+$2533.  J1626$+$2533 does not show consistent pulsation modes and J0634$-$3848 has both long and short modes. J0912$-$2642 potentially has low amplitude modes that were below our detection limits. Additional observations of these white dwarfs will help constrain their nature. 

We find a correlation between the weighted mean period and stellar mass. Theory predicts a decreasing period with increasing stellar mass, which
we confirm. This is due to higher mass white dwarfs having a higher surface gravity, leading to larger Brunt-V\"{a}is\"{a}l\"{a} frequencies and shorter
pulsation periods. Even though crystallization restricts $g$-mode propagation and should lead to lower pulsation frequencies, its effect is smaller
than the increasing Brunt-V\"{a}is\"{a}l\"{a} frequencies with mass. This is the first time a large sample of pulsating white dwarfs is found to show this relationship between weighted mean period and stellar mass. 

Pulsation periods and power mostly follow expectations. \citet{mukadam} found power to increase as temperature decreases, peaking in the middle of the instability strip, then decreasing again towards the red edge of the strip. Our sample supports this trend as the white dwarfs in our sample found in the middle of the strip have the highest power. Additionally, our sample shows a decrease in power towards the red edge of the strip. On the blue edge, J0912$-$2642 and J1626$+$2533 have the lowest power in our sample which is to be expected, but these two stars are outliers in our sample and require more observation. 

Half of the white dwarfs we observed showed no pulsation modes despite falling within the bounds of the instability strip. This invites the question, why do we not detect pulsations in these white dwarfs? The best explanations for these phenomena are magnetism, low-amplitude variability below the detection threshold, and discrepancies in temperature estimates through the photometric and spectroscopic methods. Because magnetism could be
suppressing pulsations, it is possible that the non-variables we find in the middle of the strip are weakly magnetic \citep{tremblay,gentile} with fields beyond the detection limit of low resolution spectroscopy \citep{kilic2015, jewett2024}.
Our observations of J0959$-$1828 using ULTRACAM and HIPERCAM show that there are low-amplitude modes that we simply cannot
detect with our current setup at APO, meaning more observations are required for the NOVs in the middle of the strip. 

Our study presents a more complete picture of the ZZ Ceti instability strip for massive and ultra-massive white dwarfs in the MWDD 100 pc sample. As massive and ultra-massive ZZ Ceti white dwarf are rare, these observations provide valuable insight on the most massive pulsators and trends within this class of white
dwarfs. We plan on obtaining extensive time-series photometry of the most massive systems in this sample at APO, NTT/ULTRACAM, and GTC/HiPERCAM for detailed
asteroseismology.  

\begin{acknowledgements}

This work is supported in part by the NSF under grant  AST-2508429, and the NASA under grants 80NSSC22K0479, 80NSSC24K0380, and 80NSSC24K0436.
MU gratefully acknowledges funding from the Research Foundation Flanders (FWO) by means of a junior postdoctoral fellowship (grant agreement No. 1247624N).
This work was supported by PIP 112-200801-00940 grant from CONICET, grant G149 from the University of La Plata, PIP-2971 from CONICET (Argentina) and by
PICT 2020-03316 from Agencia I+D+i (Argentina). This work was partially supported by the MINECO grant PID2023-148661NB-I00 and by the AGAUR/Generalitat
de Catalunya grant SGR-386/2021.
VSD, ULTRACAM and HiPERCAM are supported by STFC grant  ST/Z000033/1.

The Apache Point Observatory 3.5-meter telescope is owned and operated by the Astrophysical Research Consortium.

Based on observations made with the Gran Telescopio Canarias  (Prog. ID: GTC19-24B), installed at the Spanish Observatorio del Roque de los Muchachos of the Instituto de Astrofísica de Canarias, on the island of La Palma. 

Based on observations collected at the European Organisation for Astronomical Research in the Southern Hemisphere under ESO programme 0114.D-0352(C).

\end{acknowledgements}

%\begin{contribution}
%\end{contribution}

\facilities{ARC 3.5m (ARCTIC), GTC (HiPERCAM), NTT (ULTRACAM)}
\software{Period04 \citep{lenz}}
        
\bibliography{references.bib}{}
\bibliographystyle{aasjournalv7}

\appendix

\section{Additional Pulsators}
\label{appendix:a}
\subsection{J0154$+$4700}

\begin{figure*}[h]
\centering
\includegraphics[width=0.9\linewidth]{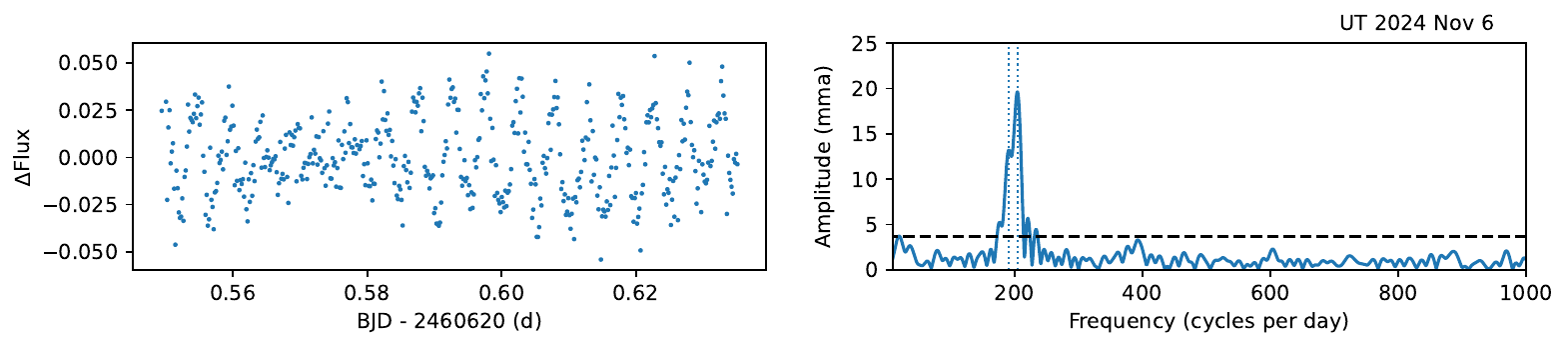}
\caption{Light curve and Fourier transform for J0154$+$4700.}
\label{fig:j0154}
\end{figure*}

J0154$+$4700 has a mass of $M=1.077 \pm 0.014~M_\odot$ and $T_{\rm eff} = 11,838 \pm 235$ K. We observed this white dwarf once on UT 2024 Nov6. 
Figure \ref{fig:j0154} shows its light curve and the corresponding Fourier transform. This white dwarf clearly shows 2 significant modes with frequencies 
190.7 and 204.9 c/d. Both of these are well over the detection limit of 3.7 mma at amplitudes 13.8 and 20.2 respectively. Our goal in this study
is to identify and confirm pulsating massive white dwarfs to study the sample properties, and we are able to confirm multi-periodic oscillations
in this object based on this data. More extensive follow-up photometry on this target, as well as the other pulsators
found in our survey would be required to characterize the pulsation spectrum of each object and to perform asteroseismology. 

\subsection{J0204$+$8713}

\begin{figure*}
\vspace{-0.2in}
\centering
\includegraphics[width=0.9\linewidth]{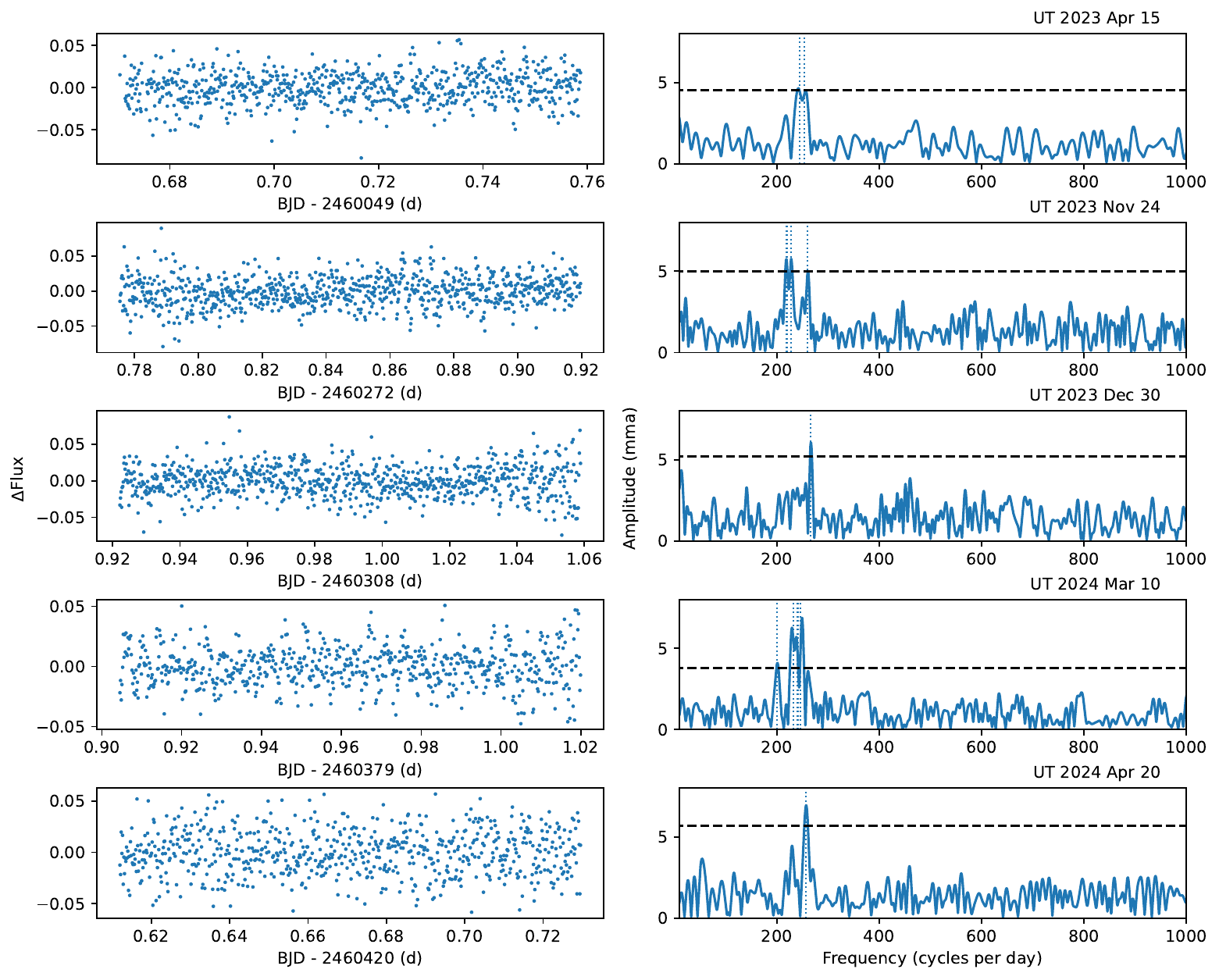}
\caption{Light curves and Fourier transforms for J0204+8713 taken at APO.} 
\label{fig:j0204}
\end{figure*}

J0204$+$8713 has a mass of $M=1.053 \pm 0.015~M_\odot$ and $T_{\rm eff} = 11,135 \pm 207$ K. This white dwarf was first identified as a pulsating white
dwarf by \citet{vincent}, where they detected a single mode at 330 s. There was no spectra available in the literature, but \citet{jewett2024} confirmed this white dwarf to be a DA and presented the first night of data from UT 2023 Apr 15 further confirming the pulsations in this white dwarf. We obtained 4 additional nights of data on this target at APO, which are shown in Figure \ref{fig:j0204}. 

\begin{figure*}
\vspace{-0.2in}
\centering
\includegraphics[width=0.9\linewidth]{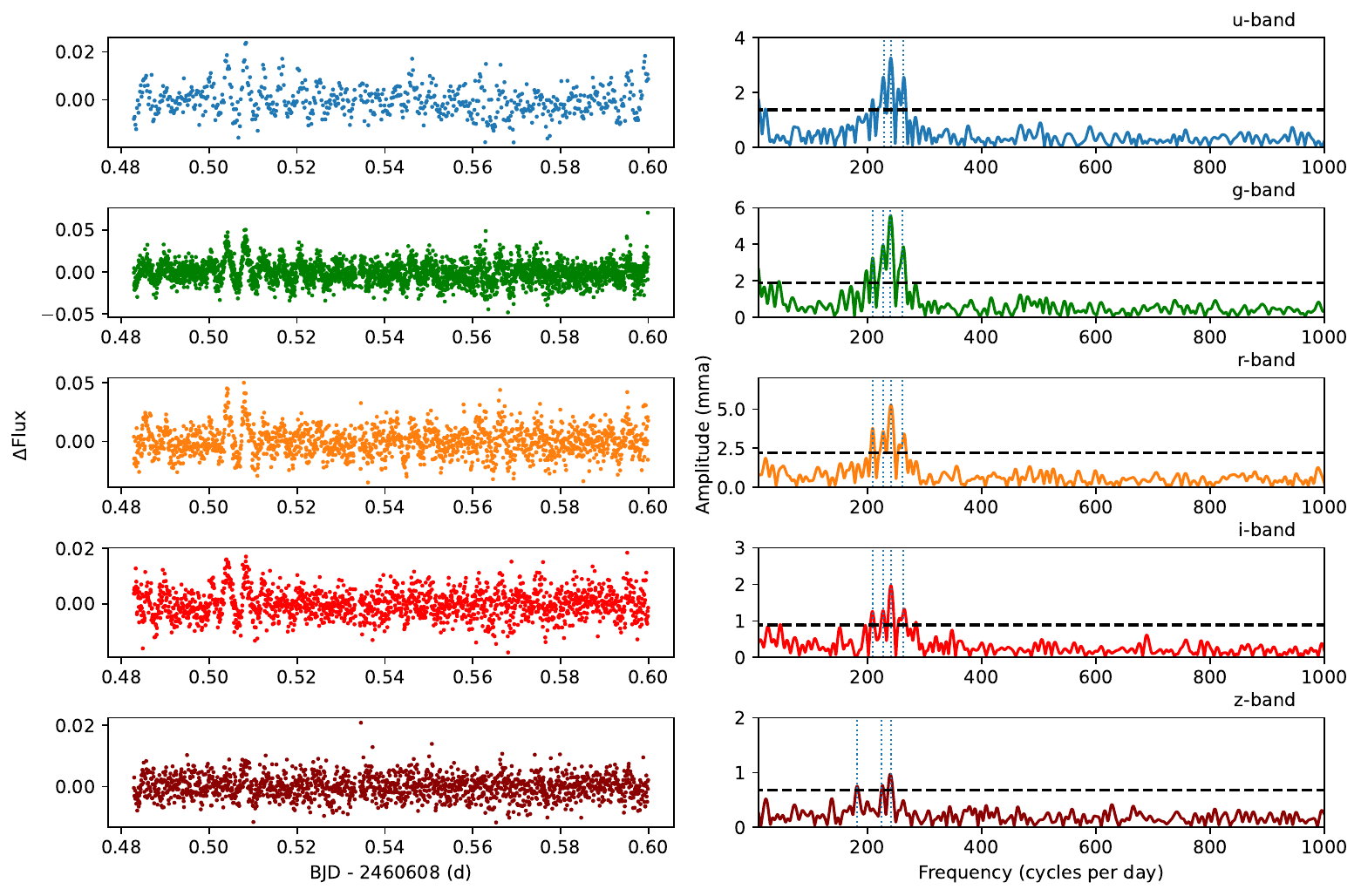}
\caption{Light curves and Fourier transforms for J0204+8713 taken at GTC.} 
\label{fig:j0204_hipercam}
\end{figure*}

Again, with this white dwarf we detect different modes on different nights. On our first night of observation, UT 2023 Apr 15, we find 2 modes with frequencies
245.0 and 253.5 c/d. The second night, UT 2023 Nov 24, we detect 3 modes with frequencies 219.4, 228.0, 260.3 c/d. On the third night, UT 2023 Dec 30, we only
detect one mode at 266.6 c/d. Then on UT 2024 Mar 10, we detect the most modes. These have frequencies of 200.6, 231.9, 240.7, and 245.8 c/d. On the fifth night, on UT
2024 Apr 20, we find a single mode at 257.4 c/d.  Finally on 2024 Oct 24, we observed J0204+8713 at GTC. Figure \ref{fig:j0204_hipercam} shows the light curves and Fourier transforms for all 5 filters used. This observation run was 3.8 hours long, but due to telescope tracking issues we removed the last 1.0 hour of data.  The $g$-band data has modes at 210.1, 228.4, 240.5, and 261.3 c/d. While we do not see the same modes
on each night, the modes we see fall in a narrow frequency range of 200-266 cycles per day (or 324-430 s). The amplitudes for the modes we detect range from 3.4$-$9.1 mma.

\subsection{J0634$+$3848}

\begin{figure*}
\vspace{-0.2in}
\centering
\includegraphics[width=0.9\linewidth]{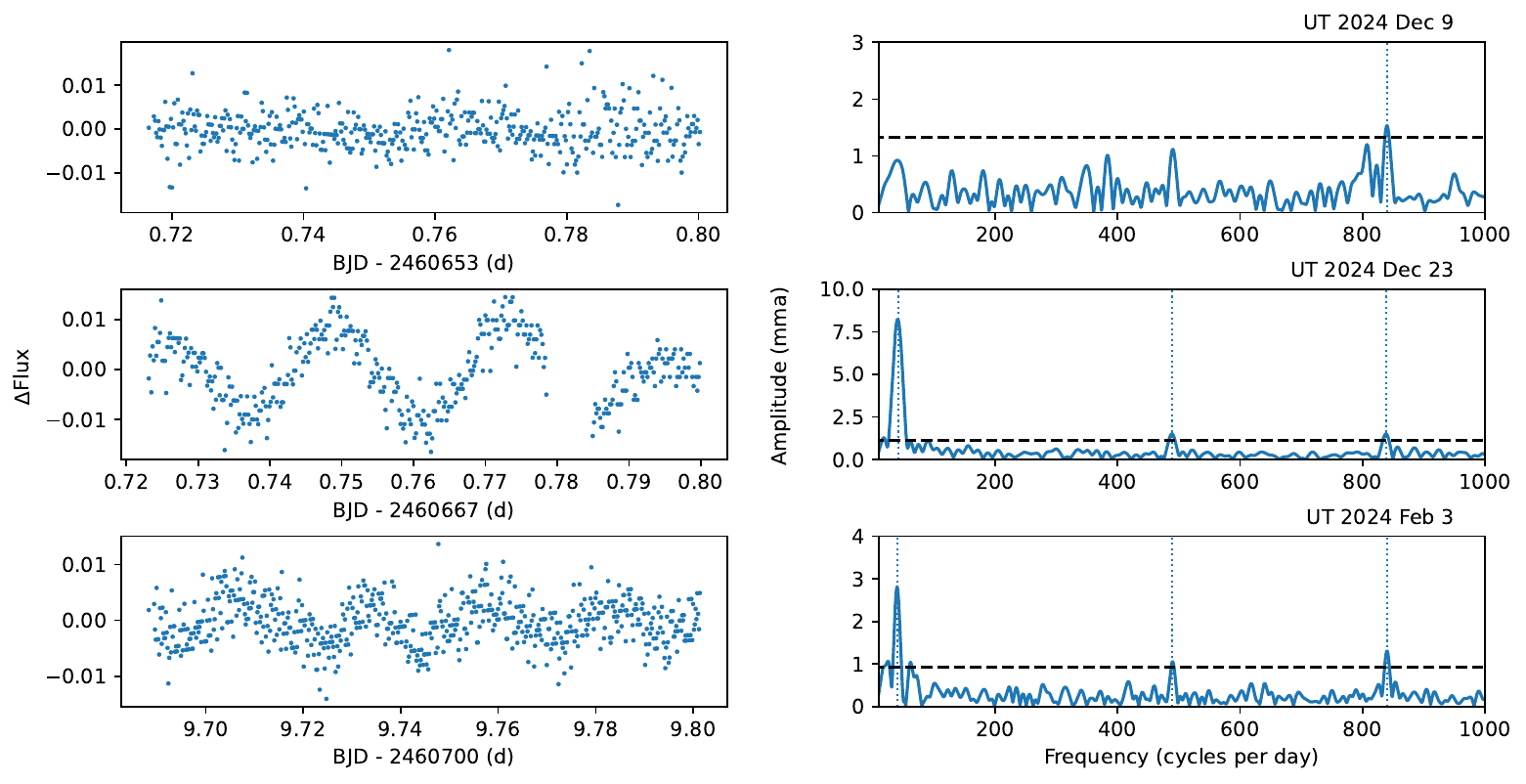}
\caption{Light curves and Fourier transforms for J0634$+$3848.}
\label{fig:j0634}
\end{figure*}

This white dwarf was previously classified as a NOV by \citet{vincent} down to a limit of 2.3\%. J0634+3848 has a mass of $M=0.926 \pm 0.006~M_\odot$ and
$T_{\rm eff} = 12,210 \pm 106$ K. We observed J0634$+$3848 on 3 different nights, and detected pulsation modes on each night giving us confidence that this
white dwarf is, in fact, pulsating. The light curves and corresponding Fourier transforms are shown in Figure \ref{fig:j0634}. 

On UT 2024 Dec 9, we find only one significant mode at 839.8 c/d. The next night of observation, UT 2024 Dec 23, we detected 3 significant modes at 41.1,
488.6, and 838.3 c/d. During the last night of observations, UT 2025 Feb 3, we also detect 3 significant modes at 40.0, 489.6, and 839.9 c/d. The
Fourier transform for J0634+3848 is clearly unusual. The mode at 103 s is clearly present in all 3 nights of data. The long period peak at $\sim$2,100 s
unusual, as it is much longer than the pulsation periods seen in average mass or massive ZZ Ceti stars. Rapid rotation in spotted white dwarfs
can cause such long term trends, but we see the 2100 s mode only on two of the nights, and its amplitude is also significantly different on those two
nights. Hence, J0634$+$3848 must be a pulsating white dwarf, but further follow-up observations are required to constrain its pulsation spectrum
and understand its nature.

\subsection{J0712$-$1815}

\begin{figure*}[h]
\centering
\includegraphics[width=0.9\linewidth]{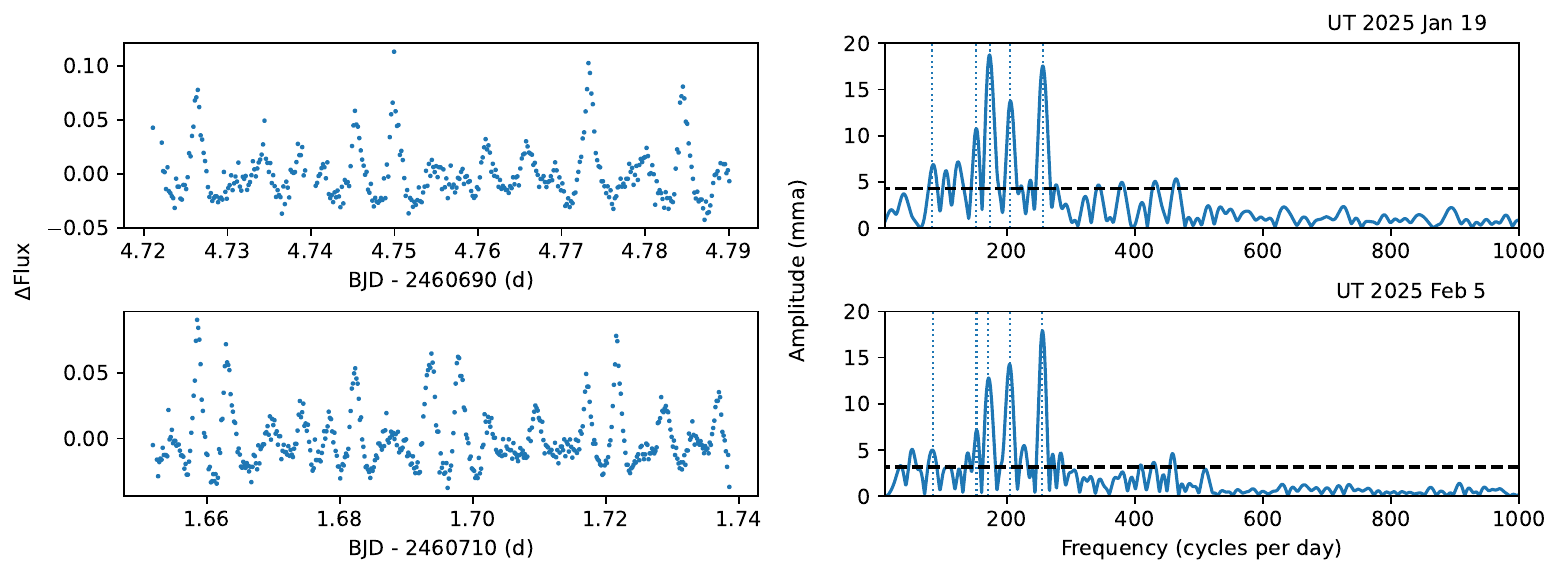}
\caption{Light curves and Fourier transforms for J0712$-$1815 }
\label{fig:j0712}
\end{figure*}

J0712$-$1815 has a mass of 0.980 $\pm$0.008 $M_\odot$ and effective temperature 11,742 $\pm$136 K. J0712$-$1815 was observed on 2 different nights. We detect 5 modes each night of observations. The first night, UT 2025 Jan 19, the modes are at 83.9, 151.7, 173.8, 205.5, and 256.9 c/d, whose amplitudes vary between 7$-$16 mma and are above the 4.3 mma significance level. We detect modes on UT 2025 Feb 5 at 85.4, 152.9, 170.7, 205.7, and 255.6 c/d, with amplitudes above the 3.2 mma significance level varying between 5$-$18 mma. The light curves and Fourier transforms are presented in Figure \ref{fig:j0712}. The light curves from these two nights look fairly similar, and we detect roughly the same modes on each of these nights making this white dwarf one of our more consistent pulsators with multiple modes detected. 

\subsection{J1052$+$1610}
\begin{figure*}[h]
\centering
\includegraphics[width=0.9\linewidth]{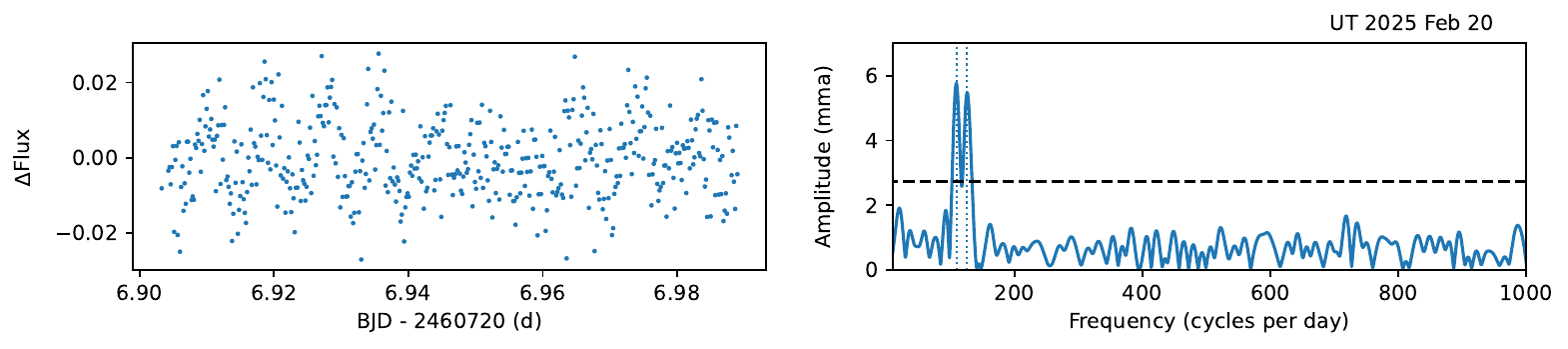}
\caption{Light curve and Fourier transform for J1052$+$1610 }
\label{fig:j1052}
\end{figure*}

J1052$+$1610 has mass 1.020 $\pm$0.009 $M_\odot$ and effective temperature 11,256 $\pm$84 K. We observed J1052$+$1610 on UT 2025 Feb 20 and detect 2 modes at 110.1 and 125.5 c/d with amplitudes above the 2.7 mma significance level at 5 mma and 4.8 mma, respectively. The light curve and Fourier transform are shown in Figure \ref{fig:j1052}.

\subsection{J1451$-$2502}
\begin{figure*}[h]
\vspace{-0.2in}
\centering
\includegraphics[width=0.9\linewidth]{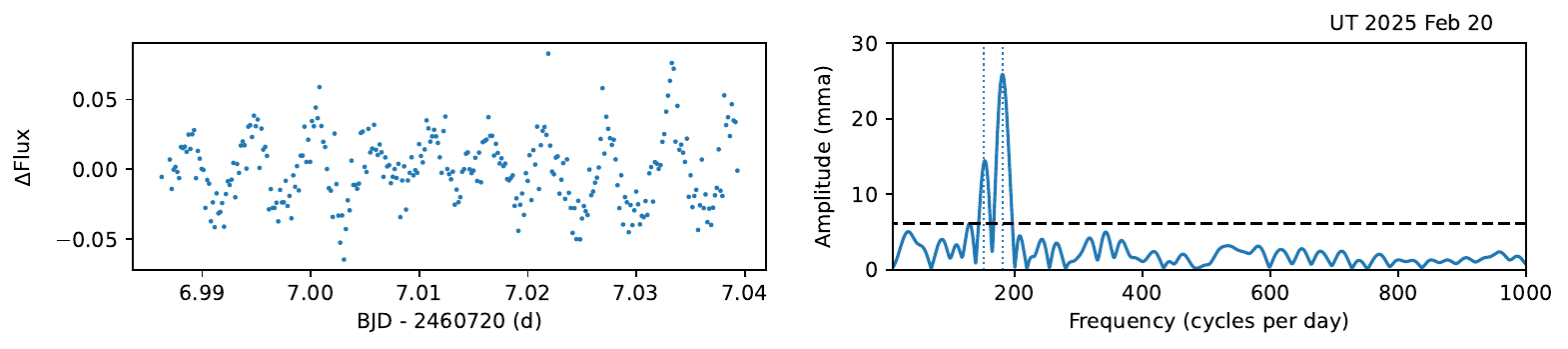}
\caption{Light curve and Fourier transform for J1451$-$2502 }
\label{fig:j1451}
\end{figure*}

J1451$-$2502 has a mass of 1.016 $\pm$0.011 $M_\odot$ and effective temperature 11,189 $\pm$97 K. We were able to observe J1451$-$2502 on UT 2025 Feb 20 and we detected 2 modes at 152.4 and 181.9 c/d, both with amplitudes over the 6.1 mma significance level. The light curve and Fourier transform are shown in Figure \ref{fig:j1451}.

\begin{figure*}
\vspace{-0.2in}
\centering
\includegraphics[width=0.9\linewidth]{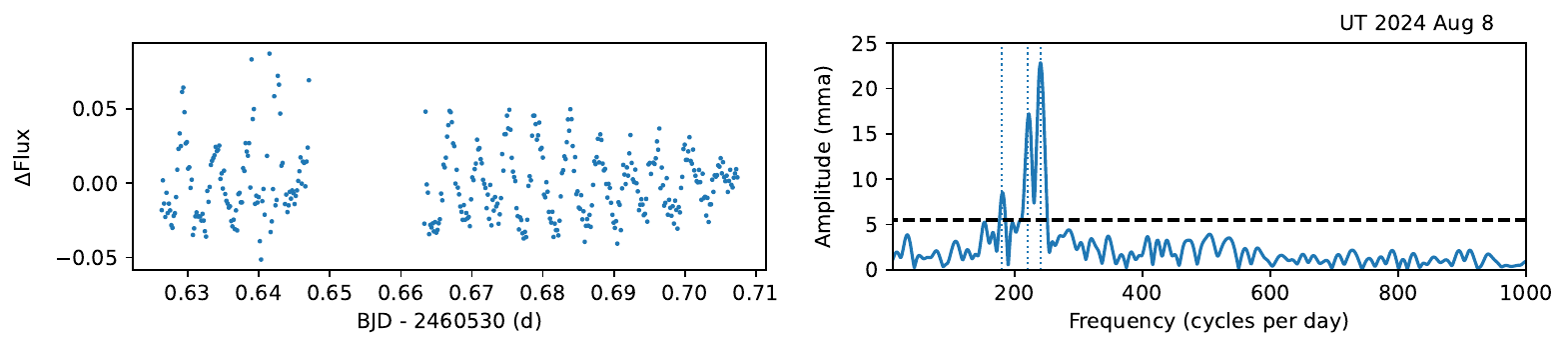}
\caption{Light curve and Fourier transform for J1722$+$3958}
\label{fig:j1722}
\end{figure*}

\begin{figure*}
\vspace{-0.2in}
\centering
\includegraphics[width=0.9\linewidth]{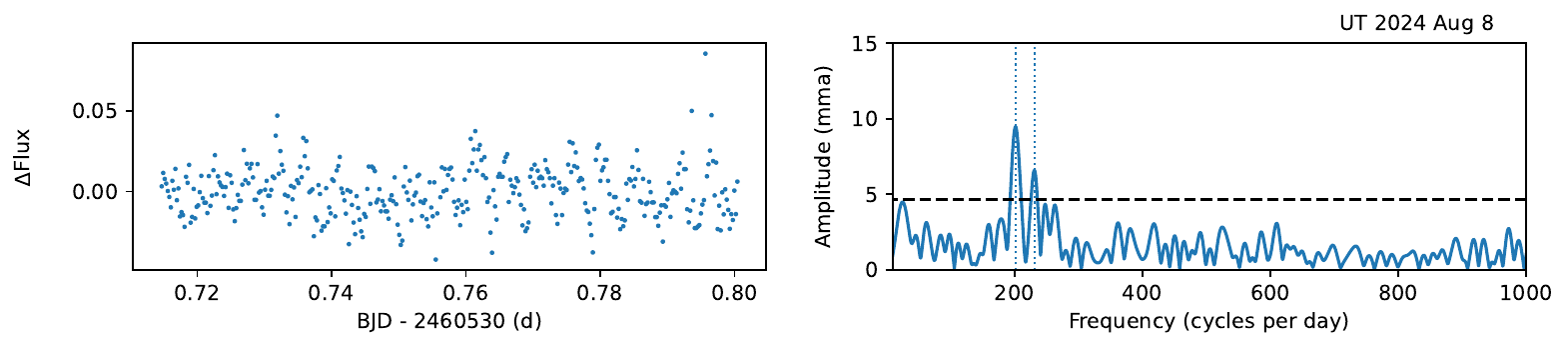}
\caption{Light curve and Fourier transform for J1929$-$2926 }
\label{fig:j1929}
\end{figure*}

\begin{figure*}
\vspace{-0.2in}
\centering
\includegraphics[width=0.9\linewidth]{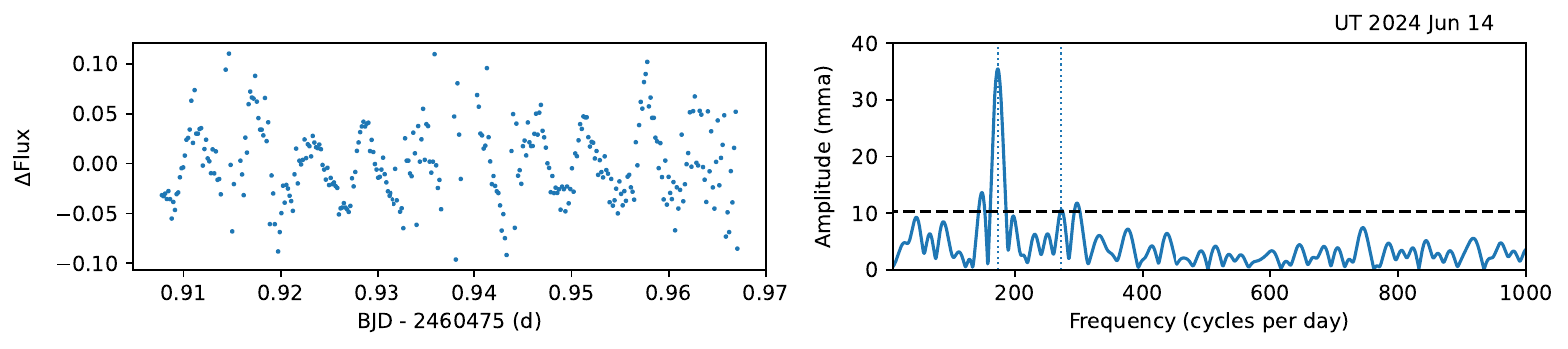}
\caption{Light curve and Fourier transform for J2026$-$2254 }
\label{fig:j2026}
\end{figure*}

\subsection{J1722$+$3958}
J1722$+$3958 has mass 0.997 $\pm$0.010 $M_\odot$ and effective temperature 11,069 $\pm$109 K. J1722$+$3959 was observed once on UT 2024 Aug 30. We find this white dwarf to have 3 modes at 180.6, 221.5, and 241.5 c/d, with amplitudes ranging from 8.1$-$19.7 mma. The light curve and Fourier transform are shown in Figure \ref{fig:j1722}.

\subsection{J1929$-$2926}
J1929$-$2926 has mass 1.054 $\pm$0.016 $M_\odot$ and effective temperature 11,455 $\pm$220 K.J1929$-$2926 was observed once on UT 2024 Aug 8. We detected 2 modes at 201.7 and 231.1 c/d with amplitudes 10.0 and 7.2, respectively. The light curve and Fourier transform are presented in Figure \ref{fig:j1929}.

\subsection{J2026$-$2254}
J2026$-$2254 has mass 0.917 $\pm$0.008 $M_\odot$ and effective temperature 11,405 $\pm$88 K. On UT 2024 Jun 14, we detect 2 modes at 174.0 and 272.1 c/d with amplitudes 35.8 mma and 11.5 mma. The light curve and Fourier transform are shown in Figure \ref{fig:j2026}. 

\subsection{J2208$+$2059}
\begin{figure*}[h]
\centering
\includegraphics[width=0.9\linewidth]{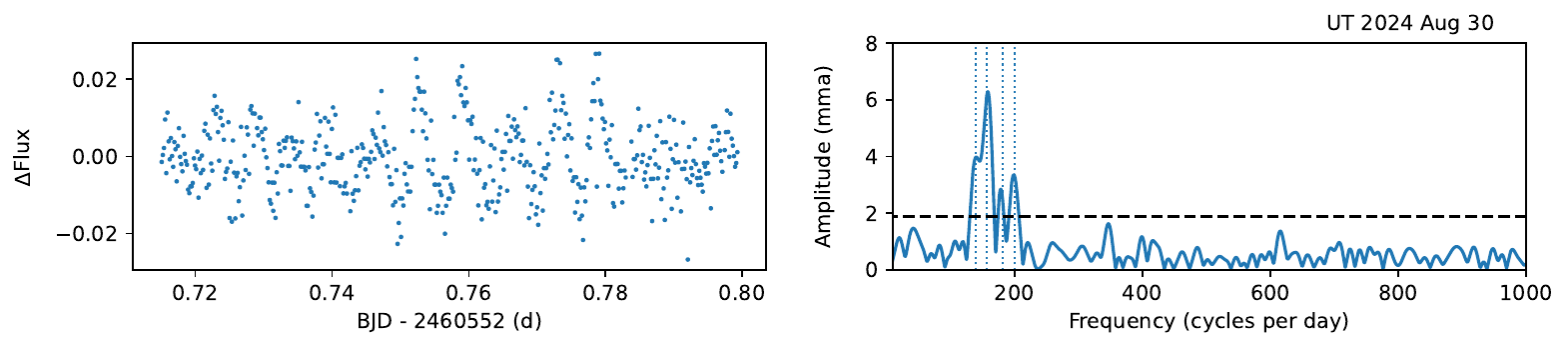}
\caption{Light curve and Fourier transform for J2208$+$2059 }
\label{fig:j2208}
\end{figure*}

J2208$+$2059 has mass 0.941 $\pm$0.011 $M_\odot$ and effective temperature 11,091 $\pm$99 K.We were able to observe this white dwarf once on UT 2024 Aug 30. We found 4 modes at 139.8, 157.5, 182.1, and 200.1 c/d with amplitudes ranging from 2.2$-$7.4 mma. The light curve and Fourier transform are shown in Figure \ref{fig:j2208}.

\section{Not-Observed-to-Vary (NOV) White Dwarfs}
\label{appendix:b}
Here we present the light curves and the corresponding Fourier transforms for 15 white dwarfs that do not show any significant variability in our follow-up
data. Each row is labeled with the object name and the observation date. The last four objects (J1656+5719, J1819+1225, J1928+1526, and J2107+7831) were
observed on two different nights, and both nights of data do not show any significant pulsation modes. The $4\langle{A}\rangle$ detection limits range from
1.5 mma in the best case (J0725+0411) to 7.7 mma in the worst case (J1928+1526), with a median at 3.1 mma. These detection limits are included in Table \ref{tab:nov}. In addition, we observed J0408+2323 and J0657+7241 at GTC using HiPERCAM. We show all 5 filters of data, with each filter labeled above the corresponding Fourier Transform. The detection limits ($4\langle{A}\rangle$) are listed in Table \ref{tab:nov_hiper}. There was also no significant variability detected in these observations that would indicate these stars are pulsating. There are two low frequency peaks in the $g$-band data in the GTC data for J0657+7241. These peaks have periods of 3223.0 s and 5162.8 s. These periods are too long to be from pulsations and are more likely from rotation. Furthermore, if this white dwarf were weakly magnetic, this could explain why it is not pulsating. We tentatively classify this star as variable due to rotation, but further observations are required to constrain these peaks. 

\begin{deluxetable}{clcc}[h]
\tabletypesize{\scriptsize}
\tablecolumns{4} \tablewidth{0pt}
\caption{Observation details and detection limits for targets that are not observed to vary (NOV) taken at APO.}
\label{tab:nov}
\tablehead{\colhead{Object Name} & \colhead{UT Date} & \colhead{Length (hr)}  & \colhead{{4$\langle {\rm A}\rangle$} (mma)} }
\startdata 
J0050$-$2826 & 2024 Dec 9 & 1.9 & 5.3\\
J0127$-$2436 & 2024 Dec 21 &1.4 & 7.2\\
J0408+2323 & 2024 Nov 21 &2.0 & 2.3\\
J0538+3212 & 2024 Dec 1 &2.1 & 1.8\\
J0657+7341 & 2024 Nov 21 & 2.0& 3.4\\
J0725+0411 & 2024 Dec 25 & 2.6& 1.5\\
J0949$-$0730 & 2025 Feb 9 &1.7 & 3.2\\
J0950$-$2841 & 2025 Mar 9 & 2.2& 6.2\\
J1243+4805 & 2024 May 18 & 2.0& 2.1\\
J1342$-$1413 & 2024 Apr 20 &1.5 & 2.0\\
J1552+0039 & 2024 Jun 6 & 2.0& 3.3\\
J1656+5719 & 2024 Aug 11 &2.6 & 1.9\\
J1656+5719& 2024 Sep 1 & 2.2& 2.2\\
J1819+1225 & 2024 Jun 14 & 1.9& 3.1\\
J1819+1225 & 2024 Sep 13 &1.5 & 5.4\\
J1928+1526 & 2024 Aug 29 & 1.3& 2.8\\
J1928+1526 & 2024 Sep 15 & 1.7& 7.7\\
J2107+7831 & 2024 Aug 29 & 2.0& 2.4\\
J2107+7831 & 2024 Sep 13 & 1.1& 4.4\\
\enddata
\end{deluxetable}

\begin{deluxetable}{ccccc}[h]
\tabletypesize{\scriptsize}
\tablecolumns{4} \tablewidth{0pt}
\caption{Observation details and detection limits for targets that are not observed to vary (NOV) taken at GTC.}
\label{tab:nov_hiper}
\tablehead{\colhead{Object Name} & \colhead{UT Date} & \colhead{Length (hr)}  & \colhead{{4$\langle {\rm A}\rangle$} (mma)} }
\startdata 
J0408$+$2323 $u$-band & 2025 Jan 16 & 2.7 &  0.4\\
J0408$+$2323 $g$-band & &  &  0.5\\
J0408$+$2323 $r$-band & &  &  0.2\\
J0408$+$2323 $i$-band & &  &  0.2\\
J0408$+$2323 $z$-band & &  &  0.2\\
\hline 
J0657$+$7341 $u$-band & 2025 Jan 10 & 4.0 &  3.2\\
J0657$+$7341 $g$-band & &  &  0.5\\
J0657$+$7341 $r$-band & &  &  0.5\\
J0657$+$7341 $i$-band & &  &  0.4\\
J0657$+$7341 $z$-band & &  &  0.4\\
\enddata
\end{deluxetable}

\begin{figure*}
\includegraphics[width=0.9\linewidth, height=1.32\linewidth]{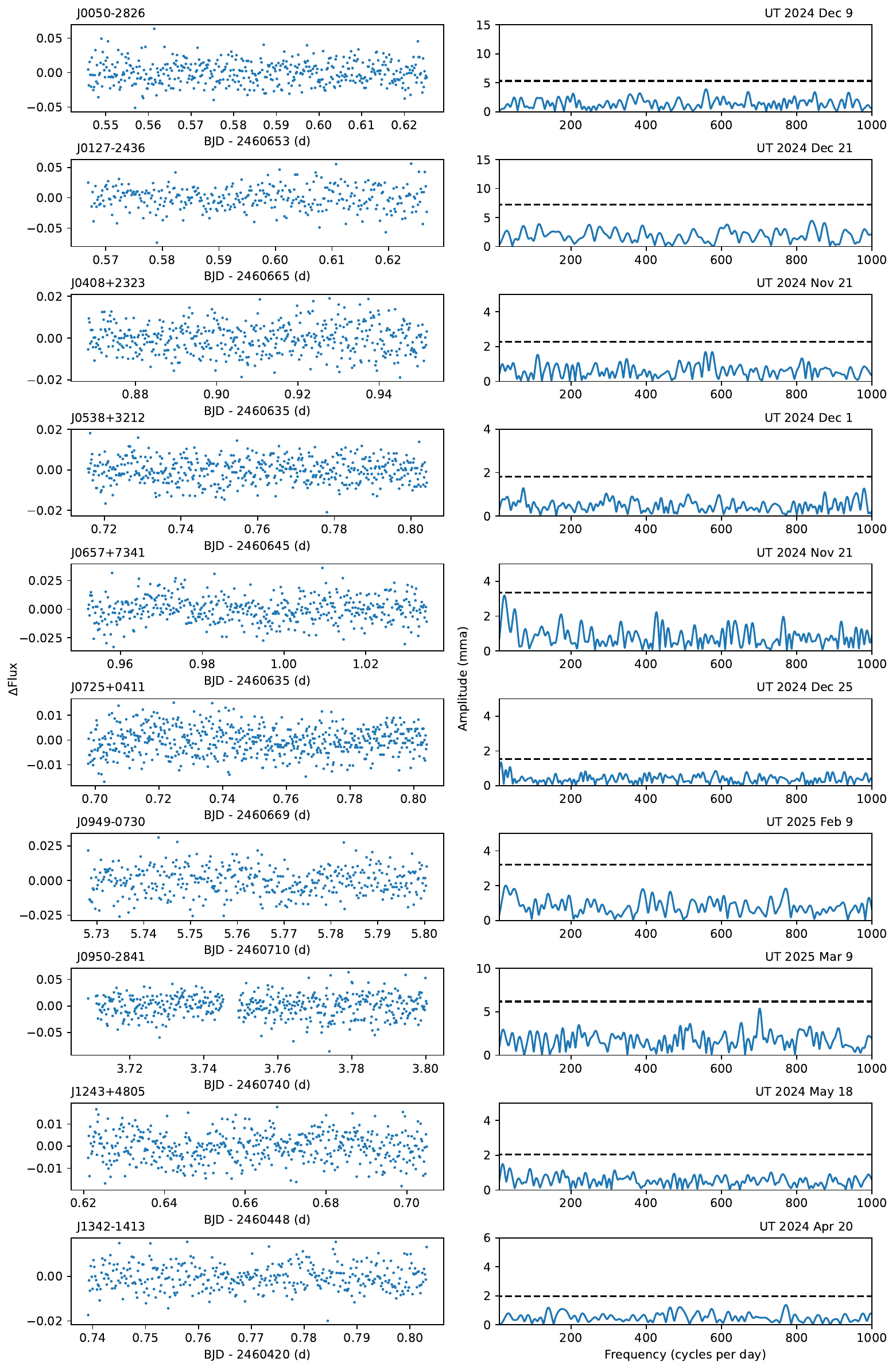}
\caption{The left panels show the light curves for non-variable white dwarfs in our sample. The right panels show the corresponding Fourier transforms with the observation date labeled on top. The dashed black lines mark the $4\langle{A}\rangle$ detection limits, which are listed in Table \ref{tab:nov}.}
\label{fig:nov}
\end{figure*} 

\addtocounter{figure}{-1}
\begin{figure*}
\includegraphics[width=0.9\linewidth, height=1.32\linewidth]{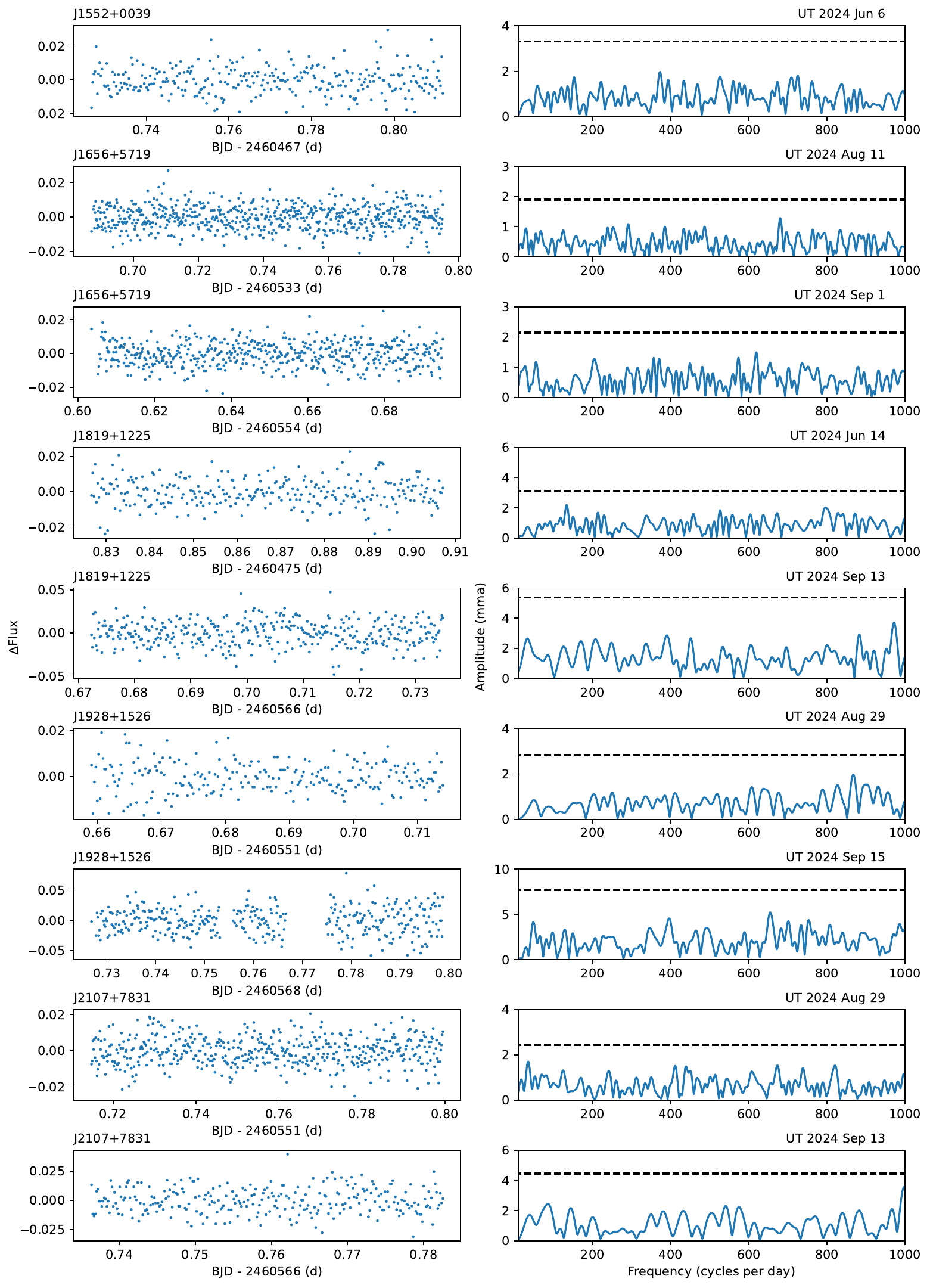}
\caption{Continued.}
\end{figure*}

\begin{figure*}
\vspace{-0.2in}
\includegraphics[width=0.9\linewidth]{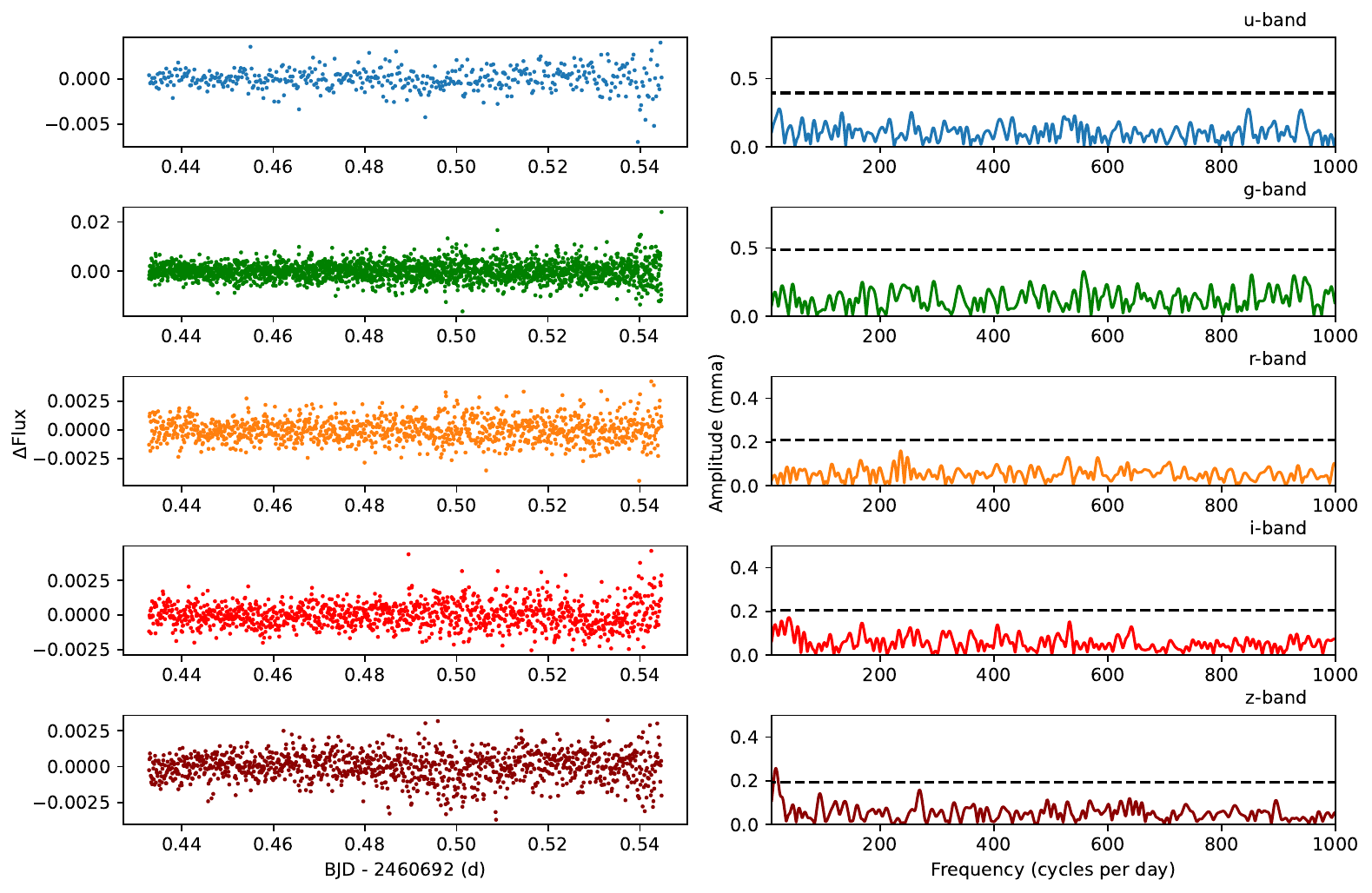}
\caption{Light curves and Fourier transforms for J0408+2323}
\label{fig:0408}
\end{figure*} 

\begin{figure*}
\vspace{-0.2in}
\includegraphics[width=0.9\linewidth]{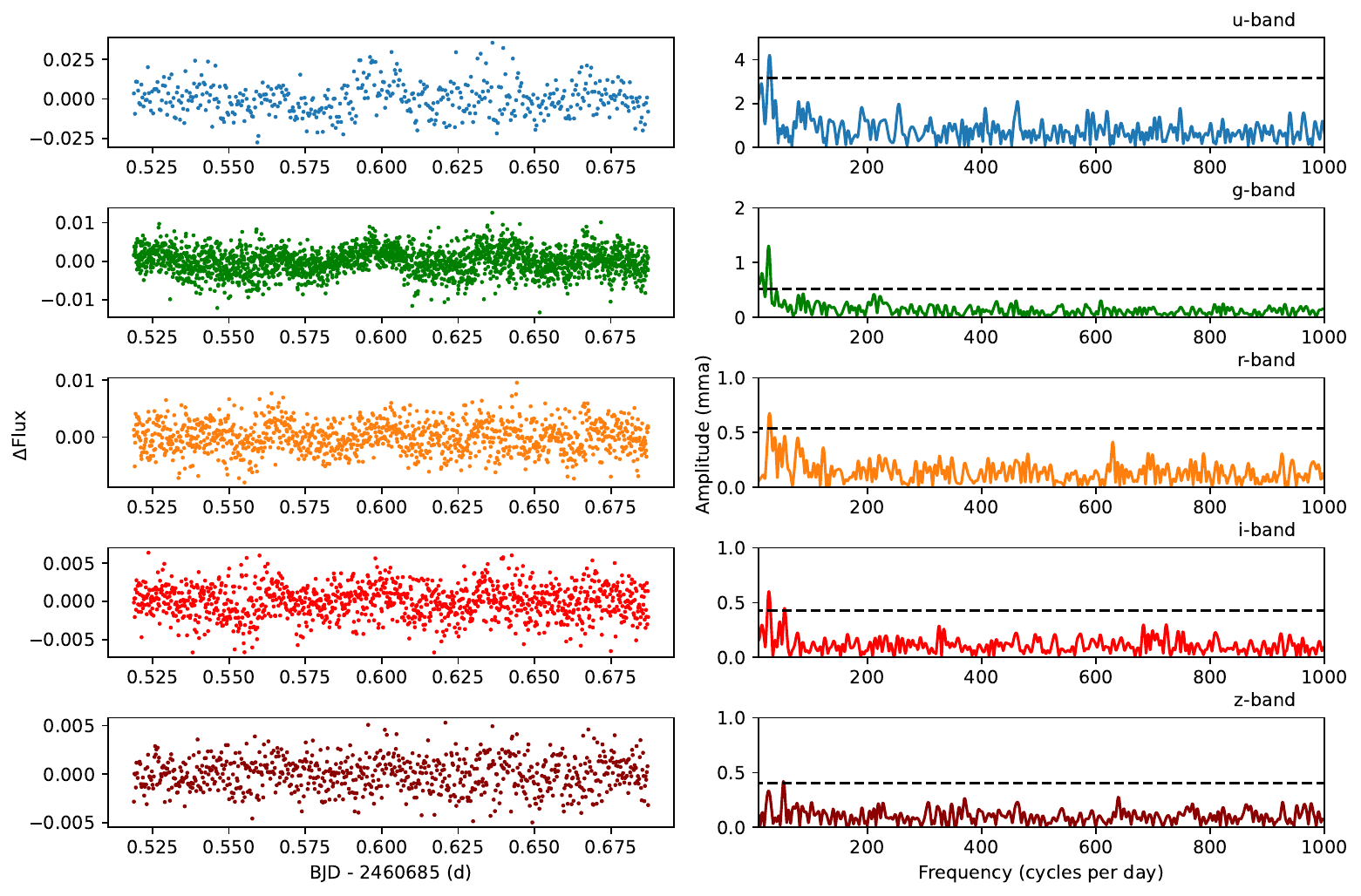}
\caption{Light curves and Fourier transforms for J0657+7341}
\label{fig:0657}
\end{figure*} 

%\begin{figure*}
%    \centering
%    \begin{subfigure}[b]{0.32\textwidth}
%        \centering
%        \includegraphics[width=\textwidth]{0050_spec.pdf}
%        \caption{J0050$-$2826}
%        \label{fig:panel1}
%    \end{subfigure}
%    \hfill
%    \begin{subfigure}[b]{0.32\textwidth}
%        \centering
%        \includegraphics[width=\textwidth]{0127_spec .pdf}
%        \caption{J0127$-$2436}
%        \label{fig:panel2}
%    \end{subfigure}
%    \hfill
%    \begin{subfigure}[b]{0.32\textwidth}
%        \centering
%        \includegraphics[width=\textwidth]{0950_spec.pdf}
%        \caption{J0950$-$2841}
%        \label{fig:panel3}
%    \end{subfigure}
%    \vskip\baselineskip
%    \begin{subfigure}[b]{0.32\textwidth}
%        \centering
%        \includegraphics[width=\textwidth]{1552_spec.pdf}
%        \caption{J1552+0039}
%        \label{fig:panel4}
%    \end{subfigure}
%    \begin{subfigure}[b]{0.32\textwidth}
%        \centering
%        \includegraphics[width=\textwidth]{1819_spec.pdf}
%        \caption{J1819+1225}
%        \label{fig:panel5}
%    \end{subfigure}
%    \caption{Model fits to five NOV white dwarfs using the spectroscopic method. Each panel shows the Balmer lines from the observed spectrum in black and the predicted spectrum using a pure hydrogen solution in red. The effective temperature and $\logg$ are shown on the bottom of the panels.}
%    \label{fig:spec fits}
%\end{figure*}

\end{document}